\documentclass{aa}

\usepackage{graphicx}
\usepackage[dvipsnames]{xcolor}

\usepackage{txfonts}

\usepackage[colorlinks=true, allcolors=blue]{hyperref}

\usepackage{placeins}
\usepackage{wrapfig}

\begin{document}

   \title{Tracing grain growth in the forming prestellar core L1506C with 3D modeling of {\it Herschel}, IRAM, and CFHT observations}

   \author{E. Zhu
          \inst{1}
          \and
        I. Ristorcelli \inst{1}  \and K. Demyk \inst{1} \and M. Juvela \inst{2} \and N. Ysard \inst{1,3} \and D. Paradis \inst{1} \and H. Roussel \inst{4} \and W. Kiviaho \inst{5}
          }

   \institute{Institut de Recherche en Astrophysique et Plan\'etologie (IRAP), Universit\'e de Toulouse, CNRS, CNES, 9 Avenue du Colonel Roche, F-31028 Toulouse, France \\
              \email{enyi.zhu@utoulouse.fr}
         \and
             Department of Physics, PO Box 64, University of Helsinki, 00014 Helsinki, Finland
         \and
         Université Paris-Saclay, CNRS, Institut d’Astrophysique Spatiale, 91405, Orsay, France
         \and
         Institut d'Astrophysique de Paris, Sorbonne Université, CNRS (UMR7095), 75014, Paris, France
         \and
         LESIA (UMR 8109), Observatoire de Paris, PSL, CNRS, UPMC, Univ. Paris-Diderot, 5 place Jules Janssen, Meudon, France
             }

   \date{Received September 15, 1996; accepted March 16, 1997}

  \abstract
   {In the early phases of star formation, dense and cold structures such as filaments and prestellar cores are commonly probed and studied via observations of thermal dust emission. The inferred core properties from these observations are critically dependent on the adopted dust model, which must be characterized carefully. However, analysis of dust properties in individual prestellar cores is often limited by the wavelength range and spatial resolution of the observations.}
   {Our target in this work, L1506C, is part of the filament L1506 in the Taurus molecular cloud ($D\sim 140$~pc). Previous studies using dust and gas tracers suggest that it is a prestellar core in the making and have revealed signatures of dust evolution. Our goal was to bring further constraints on both the cloud structures and dust properties. To do so, we used observations of dust emission covering, for the first time, a broad wavelength range, from the far-infrared (FIR) to millimeter range, as well as 3D radiative transfer modeling while also taking into account constraints from extinction data.}
   {We used both \textit{Herschel} FIR observations and complementary millimeter wavelength observations with NIKA2 at the IRAM 30m telescope. Analysis of the spectral energy distributions over the whole spectral range was first performed with modified black body (MBB) fits. The data were then analyzed with 3D modeling using the radiative transfer code SOC and the latest THEMIS 2 dust models. As additional constraints we used extinction observations from WIRCam at CFHT and from {\it Spitzer}.}
   {The MBB modeling over the wavelength range of 160~$\mu$m to 2~mm revealed that L1506C is fragmented into two low density cores with $n_\mathrm{H_2} < 5\times 10^{4} \; \mathrm{cm}^{-3}$, independently of the assumed dust opacity, with masses smaller than their Jeans masses. The dust color temperature ($T$) drops from 16~K in the filament to 11~K in the cores, and the spectral index $\beta$ increases from 1.4 to 1.9, showing a clear $T-\beta$ anticorrelation and a change in grain properties. 
   To model the source with 3D radiative transfer modeling, grains more evolved than the diffuse interstellar medium (ISM) are required in the densest part. Using grains with sizes up to$~0.7\;\mu$m, 50\% porosity, and 50\% ice content, we were able to reproduce the observed millimeter emission within a 10\% residual in the central core region and the near-infrared scattering (coreshine) within $1\sigma$. Extinction data strongly constrain the dust properties, but the best results we obtained still overestimate the extinction by a factor of more than 2.5 with the current THEMIS 2 dust model. The model $N_\mathrm{H_2}$ has a maximum value of $1.3\times 10^{22} \; \mathrm{cm}^{-2}$, 1.7 times larger than the one calculated with MBB modeling, and the model central dust temperature is down to 8~K. Good fits to the observations were obtained for the transition between diffuse and evolved grains at low density, within the range $1500 \leq n_\mathrm{H} \leq 4500 \; \mathrm{cm}^{-3}$. When adopting a model with a transition threshold of $n_\mathrm{H}=3000\; \mathrm{cm}^{-3}$, evolved grains account for 50\% of the dust content along the line of sight, where $A_\mathrm{V} \approx 4$ on the observed extinction map.}
   {Grain growth can already happen at a very early stage of star formation, even before the onset of gravitational collapse. It is difficult to precisely constrain the cloud properties with 3D radiative transfer modeling when only using dust emission observations because some of the parameters are degenerate. Observations of near-infrared extinction and scattering and independent measurements from molecular line data are crucial complements to these data, providing independent constraints and breaking degeneracies in the analysis.}


    \titlerunning{Tracing grain growth in the forming prestellar core L1506C}
   \maketitle

\section{Introduction}
The mechanisms regulating the early stages of star formation are still poorly understood. To make progress, studying in detail how dense structures form across a wide range of spatial scales -- from molecular complexes to filaments, clumps, and prestellar cores -- is essential.

The development of high-sensitivity millimeter and submillimeter instruments has opened a new era in the study of the early stages of star formation, allowing us to detect and trace the structures of cold and dense clouds through dust grain emission. Observations obtained with these instruments have provided new insights into the dust emission properties, with growing evidence that dust evolves from a diffuse to a dense medium. The characterization of dust emissivity is critical since knowledge of its nature has a direct impact on the key parameters derived, such as the densities and masses of clouds and cores. Moreover, small dust grains are the seeds of larger forming bodies in the accretion disks around young stellar objects. Therefore investigating the grain growth process already at the very early stages of prestellar cores is essential. Theoretical models have shown that dust grains can form micrometer-sized aggregates in about 1 Myr in the cores \citep{Ormel2009,Silsbee2022}. Observationally, a number of studies based on both emission and extinction data have provided indirect evidence of grain growth in dense clouds. One of the main signatures is an increase of the far-infrared (FIR)-submillimeter emissivity (or mass absorption coefficient) by a typical factor of two to three compared to the standard grains in the diffuse medium \citep{Stepnik2003, planck2011-7.13, Ysard2013, Roy2013, Juvela2015a}, as predicted by calculations of aggregate optical properties \citep{Ossenkopf1994, Koehler2012, Koehler2015}. Another striking result is the so-called coreshine, the enhanced near-IR light scattering detected with \textit{Spitzer} toward a number of dense cores \citep{Steinacker2010, Pagani2010}, which implies either the presence of very large micrometer-sized dust grains \citep{Andersen2014} or a change in dust optical properties with moderately larger dust grains \citep{Ysard2016}.

In addition to the variation of dust opacity, the study of cold core or clump spectral energy distributions (SEDs) in the FIR-submillimeter wavelength range has shown an increase of the dust emissivity spectral index ($\beta$) when the temperature drops \citep{Dupac2003, Desert2008, planck2011-7.13, planck2011-7.7b, Juvela2015b}. There is growing evidence that the $T-\beta$ anticorrelation arises in part from intrinsic grain physics, as corroborated by laboratory measurements of interstellar dust analogs, including amorphous carbons \citep{Mennella1998} and amorphous silicates \citep[][]{Agladze1996, Coupeaud2011, Demyk2017}.
Moreover, observations with NIKA and NIKA2 have enabled direct measurements of the dust emissivity spectral index $\beta$ in the millimeter range. In the Taurus B213 filament, \cite{Bracco2017} derived $\beta=2.4\pm0.3$ for a prestellar core and lower values, $\beta\approx1-1.5$, toward protostellar cores, indicating possible dust evolution. More recent NIKA2 studies of starless cores in nearby filaments have reported $\beta$ values spanning 1.1$-$2.3 and showing systematic variations with column density and environment \citep{Kramer2024}.

In this context, understanding when, where, and how dust evolves between diffuse and dense regions; what physical conditions enhance (or preventing) the coagulation process; and on what timescales dust evolution occurs remain today key questions to be investigated. In this paper, we study L1506C, a dense core embedded in a long filament in the Taurus molecular cloud \citep[at distance $\sim 140$~pc,][]{Kenyon1994, Schlafly2014} exhibiting unique dust and gas properties. Analysis of its dust emission measured along a cut with PRONAOS \citep[balloon-borne experiment with multi-band photometry in the submillimeter range,][]{Stepnik2003} and later mapped with \textit{Herschel} \citep{Ysard2013} has provided some of the first direct observational evidence of a significant increase in dust opacity in the densest part compared to the surrounding diffuse filament. This variation was interpreted as resulting from the formation of fluffy aggregates by coagulation processes. Grain growth was found to occur where gas densities exceed a few $10^3\;\mathrm{cm}^{-3}$. Molecular follow-up observations performed at IRAM in $^{13}$CO, C$^{18}$O, and N$_{2}\mathrm{H}^{+}$ lines across the filament indicate that the core is large ($r = 3\times10^{4}$~au) and has a density of $n_\mathrm{H2} \le 5\times10^{4} \; \mathrm{cm}^{-3}$, which is lower than in a typical prestellar core. Despite this relatively low density, the core is undergoing contraction with a uniform radial velocity of $100\; \mathrm{ms^{-1}}$ and is likely in the process of forming a prestellar core \citep{Pagani2010}. 

An unexpectedly strong CO depletion was observed in the central part of L1506C, with a factor of 30 (lower limit), along with the narrowest linewidth ever measured for C$^{18}$O lines -- full width at half maximum (FWHM) of $ 0.15\; \mathrm{kms}^{-1}$ -- and an extremely low turbulence \citep[lines FWHM $ \lesssim 68\; \mathrm{ms}^{-1}$,][]{Pagani2010}. This CO depletion could be an indication of ice mantle formation on dust grains, which has an impact on the grain properties (especially in extinction) and on the efficiency of aggregate formation \citep{Blum2018}. 

\cite{Juvela2024} performed a large-scale modeling of the whole L1506 filament (almost 5~pc long) by fitting radiative transfer models to \textit{Herschel} observations. This study highlights that the interpretation of the observations is highly sensitive to assumptions about the internal structure of the cloud and the spectral characteristics of the incident radiation field. Under the most plausible constraints for these parameters, the data can be accounted for only if the FIR opacity of dust grains is increased by a factor of two to three relative to values typical of the diffuse interstellar medium. However, because the analysis of \cite{Juvela2024} relies on a limited portion of the FIR spectrum, it provides only weak evidence of the presence of spatial variations in dust properties, even when considering models that extend over several square degrees of a molecular cloud.

In this study, our goal is to obtain further constraints on both the cloud structures and dust properties in L1506C by using observations of dust emission covering a broad wavelength range, from the FIR to millimeter range, and 3D modeling that includes radiative transfer. 
For this purpose, we used FIR-submillimeter observations from the \textit{Herschel} satellite (i.e., $160 - 500 \; \mu$m) and our NIKA2 follow-up in the two millimeter bands at 1.15 and 2~mm. We also used constraints from the WIRCam near-infrared (NIR) extinction map and from \textit{Spitzer} coreshine observations.

In Sect.~\ref{sec:observaiton}, we describe the observations and data used in this work. In Sect.~\ref{sec:nika2_data_processing}, we describe the data reduction process of NIKA2 maps, which includes a new approach to reconstruct the extended emission. In Sect.~\ref{sec:method} we describe in detail the methods used for our analysis. Section~\ref{sec:result} presents the results of modified black body (MBB) fits to the SEDs and of the 3D radiative transfer modeling. We discuss these results in terms of different dust properties and cloud structure in Sect.~\ref{sec:discussion} and conclude in Sect.~\ref{sec:conclusion}.

\section{Observations}\label{sec:observaiton}

\subsection{NIKA2}

\begin{figure*}[htbp]
    \centering

    \includegraphics[width=17cm]{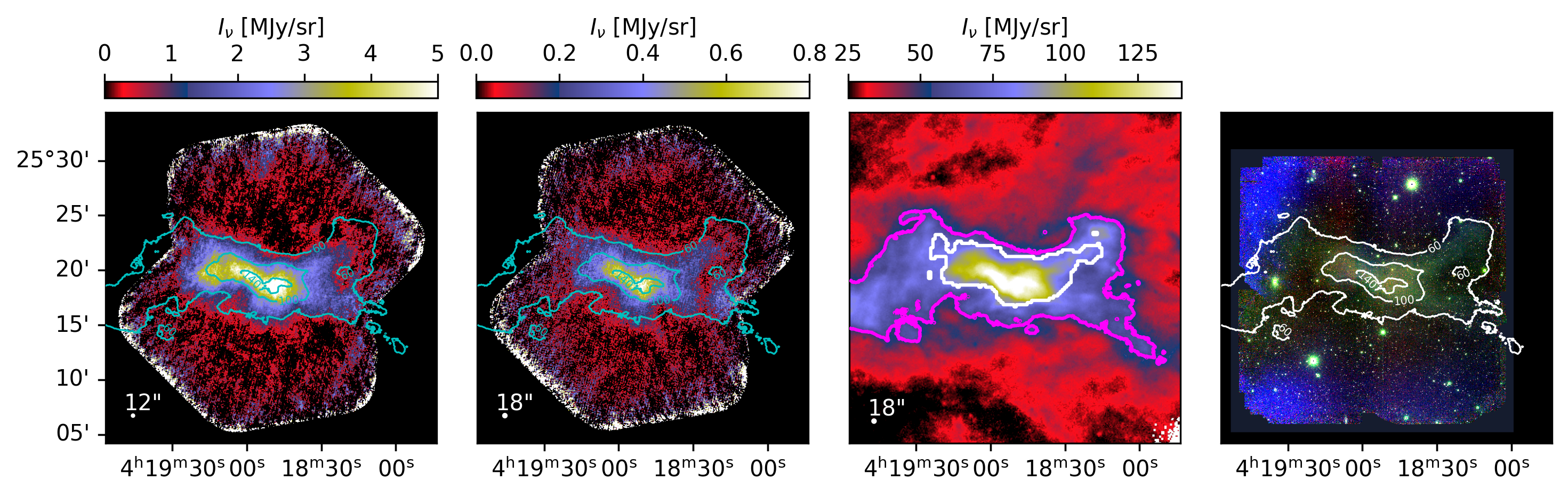}

    \caption{NIKA2 maps at 1.15~mm (leftmost panel) and 2~mm (second panel from the left) of L1506C at their nominal resolution (12$\arcsec$ for the former and 18$\arcsec$ for the latter). The cyan contours show \textit{Herschel} 250~$\mu$m intensity (at levels of 60, 100, and 140~MJy/sr). The third panel shows the \textit{Herschel} map of L1506C at 250~$\mu$m. The white contour shows the region where the MBB modeling over the whole spectral range is done (S/N$\ge 4$ for all wavelengths), and the magenta contour shows the filament region (see Sect.~\ref{sec:results_MBB_map} and Sect.~\ref{sec:result_core_properties}). The rightmost panel shows a three-color image of L1506C built from the WIRCam \textit{J} (blue), \textit{H} (green), and \textit{K} (red) band maps. White contours show the \textit{Herschel} 250~$\mu$m intensity (at levels of 60, 100, and 140~MJy/sr). The beam size is shown at the bottom left of the NIKA2 and \textit{Herschel} images.} 
    \label{fig:NIKA2_map}
\end{figure*}

Observations of L1506C with the NIKA2 camera
\citep{Adam_2018, Calvo_2016, Bourrion_2016} were taken at the IRAM 30-meter telescope in October 2020 and January 2021 as part of the project ``NIKA2 insights into the dust evolution in prestellar cores'' (PI: I. Ristorcelli, projects ID 007-20 and 094-20). The target was observed simultaneously at 1.15~mm and 2~mm with 43 scans alternating between two directions on the sky (45$^{\circ}$ and 100$^{\circ}$ relative to the RA axis), each with a length of $20\arcmin$ and a width of $17\arcmin$, providing a total field of view (FOV) with nominal coverage of $\sim 250$~arcmin$^2$. The scanning speed was 50 arcsec/s. Pointings, done every 75 minutes, and focus, every two hours on average, were of reasonable quality and repeated if needed.
The effective opacity at the elevation of the source ranged between 0.08 and 0.24 at 2\,mm and between 0.12 and 0.4 at 1.15\,mm, with a relatively stable atmosphere.
The 1.15~mm map has a beam FWHM of 12$\arcsec$ and the 2~mm map 18$\arcsec$.
The total integration time was 6.4 hours on-source. Figure~\ref{fig:NIKA2_map} shows the NIKA2 maps at their nominal resolution.

\subsection{Herschel}

We utilized the observations of L1506C at 250~$\mu$m, 350~$\mu$m and 500 $\mu$m with the \textit{Herschel} spectral and photometric imaging receiver (SPIRE) instrument as well as the observations at 160~$\mu$m with the photodetector array camera and spectrometer (PACS) instrument \citep{Griffin2010, Poglitsch2010}. These observations were carried out as part of two \textit{Herschel} guaranteed-time key projects, the \textit{Herschel} Gould Belt survey \citep[PI: P. Andr\'{e},][]{Andre2010} and Evolution of interstellar dust \citep[PI: A. Abergel,][]{Abergel2010} with a map area of 35$\arcmin \times33\arcmin$. The original beam FWHM is 12$\arcsec$, 18$\arcsec$, 25$\arcsec$ and 36$\arcsec$ respectively for wavelengths in ascending order. Figure~\ref{fig:NIKA2_map} shows the \textit{Herschel} 250~$\mu$m map of L1506C.

\subsection{WIRCam}

Apart from the \textit{Herschel} and NIKA2 data,
we used observations of stellar photometry and extended emission obtained from the Canada–France–Hawaii Telescope (CFHT) Wide-field Infrared Camera \citep[WIRCam,][]{Puget2004} in the \textit{J}, \textit{H}, and \textit{K} NIR wave bands (1.25~$\mu$m, 1.63~$\mu$m, and 2.15~$\mu$m, respectively). The observations of L1506C were obtained in 2012 under the project ``Revealing the density structure and dust properties in interstellar filaments” (P.I.: I. Ristorcelli).

\section{Data processing}\label{sec:nika2_data_processing}

\subsection{NIKA2}

\subsubsection{Default calibration and data processing}
\label{sec:scanam}

The raw data were preprocessed with the IDL pipeline developed by the NIKA2 collaboration in order to apply the opacity correction and point-source flux calibration based on dedicated, finely sampled maps of Uranus and skydip observations taken regularly during each observing run \citep{Perotto2020}. The calibrated data were then ingested into the {\it Scanam\_nika} software, to compute source-beam coupling factors for each detector that are more valid for extended sources, and to subtract the atmospheric and instrumental noise, as explained in \cite{Ejlali2025}. We also refer to the IMEGIN catalog paper (Ejlali et al. 2026, in prep.) for a thorough description. The main features of the noise subtraction are described in Appendix~\ref{appendix:nika2}. At the end of the processing, the time series were projected on a common grid at both 1.15~mm and 2~mm. When the output had to be convolved to a given FWHM larger than the native FWHM, this was done during the map projection, using a Gaussian kernel. The associated weight maps, the weight of each sample being equal to the inverse squared high-frequency noise of the corresponding detector and scan leg, were used to select the regions with nominal coverage and noise, i.e., where the weights were higher and almost uniform. We applied a conversion from Jy/beam units to surface brightness units, using the following effective beam areas, appropriate for extended sources, as explained in Appendix A.1 of \cite{Ejlali2025}: $(381 \pm 11)$~arcsec$^2$ at 2\,mm and $(188 \pm 11)$~arcsec$^2$ at 1.15\,mm.

\subsubsection{Reconstruction of large-scale emission at millimeter wavelengths}
\label{sec:data_reduction}

\begin{figure}[t]
    \centering

    \includegraphics[width=\hsize]{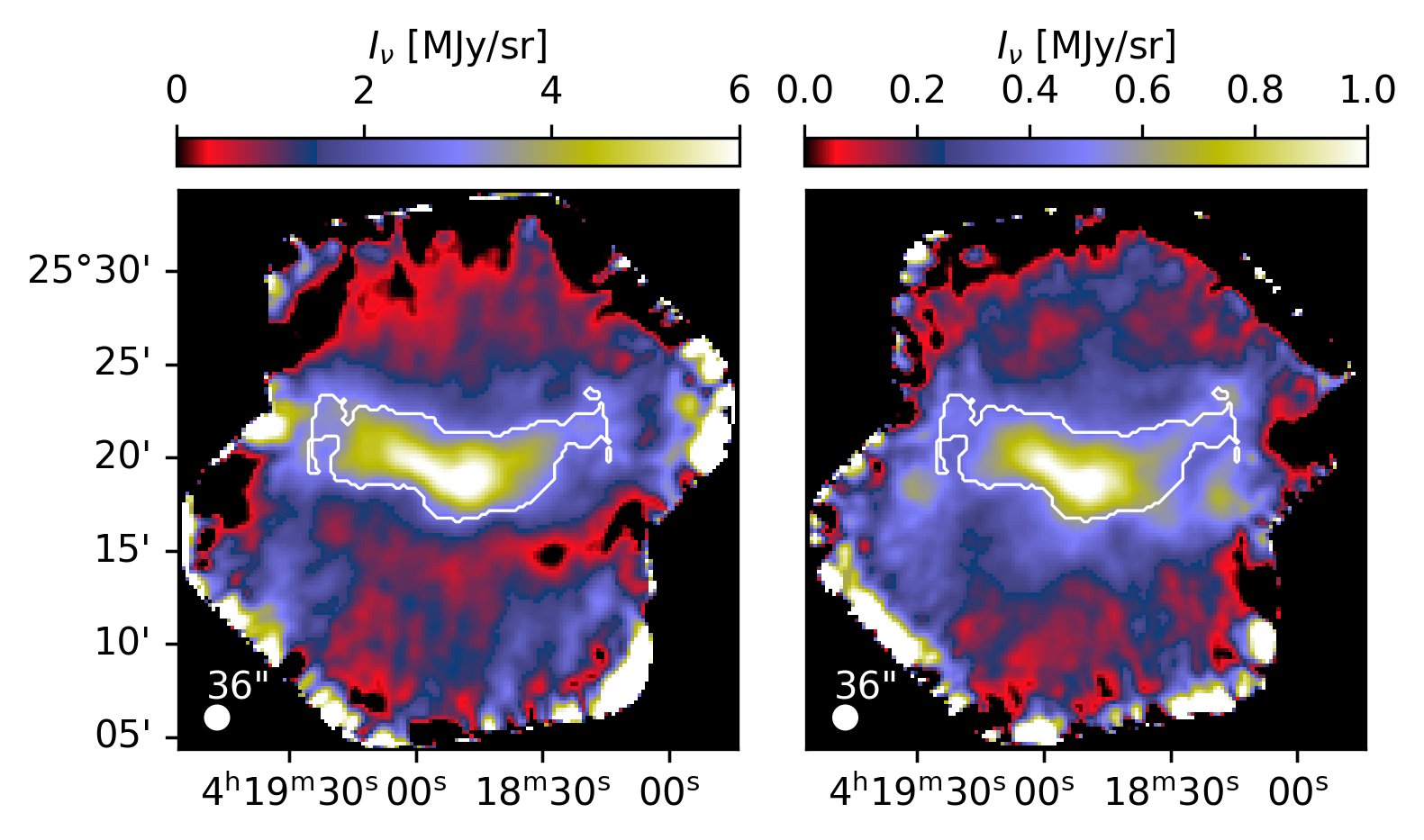}

    \caption{Reconstructed NIKA2 maps at 1.15~mm (left) and 2~mm (right) of L1506C with a pixel size of 12$\arcsec$ and a resolution of 36$\arcsec$. The beam size is shown in the bottom-left corner of each frame. The white contours show the area used for MBB modeling, where S/N$\ge 4$ in all wavelength bands.} 
    \label{fig:reconstructed_NIKA2_map}
\end{figure}

In order to combine the ground-based NIKA2 observations with the space-borne \textit{Herschel} observations, we developed an alternative data processing method. This approach provides consistent data sets over the whole spectral range, as explained below.

Compared to a space telescope, sky-noise subtraction of a ground-based telescope such as the IRAM 30m dish filters out part of the extended emission of the target mainly due to the removal of the atmospheric foreground and instrumental drifts.
The transfer function depends on the geometry, the scanning speed and depth of the observations. Even in \textit{Herschel} maps, the emission extended over scales larger than the map size is filtered out.

Conversely, the \textit{Planck} mission delivered all-sky maps of dust emission at submillimeter and millimeter wavelengths that were calibrated on an absolute scale \citep{PlanckI}. The \textit{Planck} observations are thus especially useful for the reconstruction of extended emission. L1506C is also well detected by \textit{Planck} at 850~$\mu$m and 1.38~mm. However, \textit{Planck} data suffers from a low angular resolution, which is at the highest frequencies (217~GHz to 857~GHz) of about $5\arcmin$. \cite{Paradis2024} developed a machine learning method that uses the \textit{Herschel} maps and \textit{Planck} data at 5$\arcmin$ resolution to train a neural network and predict mock observations in the \textit{Planck} bands at an angular resolution of $36\arcsec$, called the predicted maps hereafter\footnote{The prediction maps in the HEALPix format are available on the CADE website: \url{https://cade.irap.omp.eu}.}. Predictions for the NIKA2 1.15 and 2\,mm bands can also be obtained by interpolation or extrapolation using the slope between the predictions in the \textit{Planck} bands at 850~$\mu$m and 1.38~mm. 
We used these predicted maps indirectly in the data processing to recover the extended emission over scales beyond the NIKA2 array size ($6.5\arcmin$), as explained in Appendix~\ref{appendix:nika2}.
The resulting maps at 1.15 and 2 mm are called ``reconstructed maps'' hereafter.

The nominal resolutions of the NIKA2 maps are preserved during reconstruction. The reconstructed maps are shown in Fig.~\ref{fig:comparison_factor} (middle column of $f=1$). For the MBB analysis (Sect.~\ref{Sec: method_maps} and~\ref{sec:method_core_property}), all the original and reconstructed maps were smoothed to the same beam FWHM of $36\arcsec$, which was limited by the \textit{Herschel} 500~$\mu$m band that has the lowest resolution. They were also resampled to a pixel size of $12\arcsec$ to avoid oversampling. The reconstructed, smoothed and resampled NIKA2 images are shown in Fig.~\ref{fig:reconstructed_NIKA2_map}. Compared to the original NIKA2 maps (Fig.~\ref{fig:NIKA2_map}), the reconstructed maps have higher intensity, especially in the center and along the filament-like structure northwest to the central part. The peak emission is increased by $\sim20$\% at 1.15~mm and $\sim40$\% at 2~mm. Only the central part selected by S/N and/or weight is considered for the data analysis (see Sect.~\ref{Sec: method_maps}).

\subsection{WIRCam}\label{sec:wircam}

WIRCam data reduction was performed at the TERAPIX astronomical data reduction center at the Institut d’Astrophysique de Paris, using a specific reduction method to recover the extended emission. Figure~\ref{fig:NIKA2_map} shows the three-color image built from the \textit{JHK} band maps with a map area of 26$\arcmin \times26\arcmin$. 
We derived the corresponding \textit{J} band extinction ($A_\mathrm{J}$) map with a resolution of 1$\arcmin$ from the JHK band images with the NICER method \citep{Lombardi2001}. For NICER, we tried to use for each line of sight (LoS) the correct mixture of dusts, as included in the SOC modeling (see Sect.~\ref{sec:dust_model} and~\ref{sec:method_SOC}).

\subsection{CO contamination}

The $^{12}$CO(2-1) transition line at 230.538~GHz falls within the spectral range of the NIKA2 1~mm filter \citep[see e.g., Fig.~2 of][]{Bracco2017} and its possible contamination on the emission in the 1.15~mm band needs to be quantified. Following the procedure described in \cite{Rigby2018}, we used $^{13}$CO(2-1) emission of L1506C measured by \cite{Pagani2010} to estimate the $^{12}$CO(2-1) emission, for which we made assumptions of $^{13}$CO(2-1) optical depth to estimate the $^{12}$CO(2-1) optical depth. Then we converted the velocity integrated main-beam temperature to MJy/sr using Eq.~8 in \cite{Drabek2012}. The NIKA2 filter transmission curve is adopted from the IRAM NIKA2 website\footnote{\url{https://publicwiki.iram.es/SpectralBands}} and the Spanish VO filter service\footnote{\url{https://svo2.cab.inta-csic.es/svo/theory/fps/index.php?mode=browse&gname=IRAM&gname2=NIKA2&asttype=}}.
We confirm that CO contamination of L1506C is at most 5\%. Although no corrections were made, the results presented in the following remain unaffected.

\section{Method}\label{sec:method}

The \textit{Herschel} 250~$\mu$m intensity ($I_{250}$) contour (Fig.~\ref{fig:NIKA2_map}) reveals two intensity peaks at the central part that could correspond to fragmentation into two cores. We expect the cores to be embedded in an intermediate envelope within the surrounding filament. These structures are defined in detail according to the $N_{\mathrm{H}_2}$ map and $I_{250}$ map in Sect.~\ref{sec:result_core_properties}. In this section, we introduce how we study the cores and filament properties with MBB modeling applied either to the maps pixel by pixel (Sect.~\ref{Sec: method_maps}) or to the SEDs of the structures obtained with aperture photometry (Sect.~\ref{sec:method_core_property}). The structures were studied afterward with dust models (Sect.~\ref{sec:dust_model}) and 3D radiative transfer modeling (Sect.~\ref{sec:method_SOC}).

\subsection{Background subtraction}\label{sec:bkg}

A constant background level was determined and subtracted separately for the original \textit{Herschel} and reconstructed NIKA2 maps in each band. We selected areas where both $I_{250}<30$~MJy/sr and the weight of the 1.15~mm intensity map is larger than 3000 to exclude the noisy edges on the millimeter maps (Fig.~\ref{fig:bkg_mask}). A Gaussian function was fitted to the histogram of the pixel intensities within that region for each wavelength. The histogram center was considered the background value and the standard deviation as its uncertainty. For MBB modeling, aperture photometry and radiative transfer modeling presented in the following sections, the maps used are all background-subtracted.

\subsection{MBB modeling} \label{Sec: method_maps}

Under the assumption of an optically thin medium, the dust emission at FIR and submillimeter frequencies $\nu$ can be approximated by the MBB function of optically thin emission:
\begin{equation}
    I_\nu = \tau_\nu B_\nu (T) = \kappa_\nu \Sigma B_\nu (T) \propto \nu^\beta B_\nu (T) \; ,
\end{equation}
where $\tau_\nu$ is the optical depth, $B_\nu$ is the blackbody radiation, $T$ is the dust color temperature, $\kappa_\nu = \kappa_0 (\nu/\nu_0)^{\beta}$ is the dust mass absorption coefficient or the opacity, $\Sigma=m_\mathrm{H_2}\times N_\mathrm{H_2}$ is the mass surface density and $\beta$ is the dust emissivity spectral index.

The $\mathrm{H}_2$ column density map is then calculated with the following equation:
\begin{equation}
    N_{\mathrm{H}_2} = \frac{I_\nu}{\mu m_\mathrm{H}\kappa_0 (\nu/\nu_0)^{\beta} B_\nu(T)} \; ,
    \label{eq:N_H}
\end{equation}
where $\mu = 2.8$ is the molecular weight per hydrogen molecule in the molecular clouds \citep{Kauffmann2008}, $m_\mathrm{H}$ is the mass of one hydrogen atom, $\nu$ is the frequency at 250~$\mu$m and $T$ and $\beta$ are values from the MBB modeling on each pixel SED. The reference opacity, $\kappa_0 = 0.1 \,\mathrm{cm^{2}\,g^{-1}}$, is adopted according to the model of \cite{Beckwith1990} at $\nu_0 = 10^{12}$~Hz (300~$\mu$m, B90 hereafter). The wavelength of 250~$\mu$m was chosen because it is where the intensity peaks at and is the closest to the B90 reference wavelength so that extrapolation with the assumed $\beta$ has the lowest impact.

We fitted the SEDs derived from the original \textit{Herschel} maps and the reconstructed NIKA2 maps with the MBB function pixel by pixel to construct the $T$ and $\beta$ maps. Both $T$ and $\beta$ were set as free parameters. The spectral index derived from MBB fits to all wavelengths bands is called $\beta_\mathrm{all}$ hereafter. We only selected the pixels with S/N of the reconstructed maps $\ge 4$ in all wavelength bands, and neglected the noisy edge (Fig.~\ref{fig:reconstructed_NIKA2_map}). The uncertainties on the flux were calculated as the quadratic sum of the background uncertainties and of the calibration uncertainties. For the latter we assumed 7\% for all \textit{Herschel} maps, 8\% for the NIKA2 1.15~mm map and 6\% for the NIKA2 2~mm map \citep{Valtchanov2011, Balog2014, Perotto2020, Katsioli2023, Paradis2024}. The resulting flux uncertainty within the MBB fitted area ranges from 11\% to 25\% at 1.15~mm and from 9\% to 25\% at 2~mm (Fig.~\ref{fig:map_uncertainty}).

The fitting was done based on $\chi^2$ minimization. Because this method can lead to incorrect anticorrelation between $T$ and $\beta$ due to noise effect and flux uncertainty \citep{Shetty2009, Juvela2012a}, we tested the robustness of our results by also performing fits using the hierarchical Bayesian method. In the classical Bayesian method, the posterior distribution $P(\theta|d)\propto P(\theta)P(d|\theta)$, where $d$ is the observation data and $\theta$ are the model parameters, is derived from the prior distribution of the parameters, $P(\theta)$, and the likelihood function, $P(d|\theta)$. The hierarchical Bayesian method, on the other hand, can infer the prior distribution from the given sample with its shape assumed and reduce the $T-\beta$ degeneracy \citep{Kelly2012, Juvela2013, Lamperti2019}.

We also calculated $\beta$ between the two NIKA2 bands ($\beta_\mathrm{mm}$ hereafter) with the reconstructed maps according to the following equation:
\begin{equation}
    \beta_\mathrm{mm} = \frac{\ln(I_{\nu_1}B_{\nu_2}(T_{Herschel})/I_{\nu_2}B_{\nu_1}(T_{Herschel}))}{\ln(\nu_1/\nu_2)} \; ,
    \label{equation:beta_mm}
\end{equation}
where $\nu_1$ and $\nu_2$ are frequencies corresponding to the NIKA2 bands (1.15~mm and 2~mm respectively), and $T_{Herschel}$ is the temperature derived from the pixel-by-pixel MBB fit of the SEDs built using only the \textit{Herschel} maps from 160~$\mu$m to 500~$\mu$m, with $\beta$ as free parameter ($\beta_{Herschel}$). By using $T_{Herschel}$, we make sure that information of NIKA2 maps is only incorporated in $I_{\nu_1}/I_{\nu_2}$ in Eq.~\ref{equation:beta_mm}, so that terms involving \textit{Herschel} or NIKA2 data are independent. We checked that using $T$ derived with all wavelength bands does not change the results (see Sect.~\ref{sec:discussion_beta}).

\subsection{Core properties}\label{sec:method_core_property}

Aperture photometry was performed for the two embedded cores, the surrounding envelope and the outermost filament of L1506C (the identification of each region is described in Sect.~\ref{sec:result_core_properties}). The SED of each structure was built using the mean value of all pixel intensities within the respective area at each wavelength. To construct the SEDs of the cores, we subtracted the envelope's intensity from that of the cores. To construct the SEDs of the envelope and outer filament, we subtracted the calculated background intensity.

The SEDs of the cores, the envelope, and the filament were fitted with MBB functions (Sect.~\ref{Sec: method_maps}), and the $N_\mathrm{H_2}$ of the cores was calculated with Eq.~\ref{eq:N_H} using different $\kappa$ values (B90 or THEMIS 2 evolved grains; see Sect.~\ref{sec:dust_model}). The volume density $n_\mathrm{H_2}$ was calculated assuming that the two cores are prolate spheroids in 3D. The core mass $M$ was calculated as
\begin{equation}
    M = \mu m_\mathrm{H} \int_A N_\mathrm{H_2} \: dA \; ,
\end{equation}
where $A$ is the core surface area.

We compared the core mass to its Jeans mass:
\begin{equation}
    M_J = \frac{5k_B TR}{\mu m_\mathrm{H}G} \; ,
\end{equation}
where $k_B$ is the Boltzmann constant, $T$ is the core temperature from MBB fit and $G$ is the gravitational constant. The effective radius $R$ was adopted from the Knud Thomsen formula that describes an ellipsoid having the same surface area as a sphere with radius $R$ \citep[see also][]{Knud-Thomsen}:
\begin{equation}
    R = \sqrt{\left(\frac{(b\times b)^{1.6} + 2(a\times b)^{1.6}}{3}\right)^{1/1.6}} \; ,
\end{equation}
where $a$ and $b$ are the major and minor axis of the elliptical core.

\subsection{Dust models}\label{sec:dust_model}

We adopted the THEMIS 2 model, designed for diffuse interstellar medium (ISM) type dust \citep{Ysard2024}, for radiative transfer modeling. Similar to its precursor model, THEMIS \citep{Jones2013, Kohler2015, Ysard2016}, THEMIS 2 is based on the concept of continuous dust evolution in the diffuse interstellar medium. The THEMIS 2 model consists of three grain populations. Two populations consist of large grains of amorphous silicate and amorphous, hydrogen-rich carbon (a-C:H) with a core-mantle structure (CM), surrounded by an aromatic-rich, hydrogen-poor carbon mantle. The third population consists of small carbon grains. Contrary to the original THEMIS model in which the grains are spherical, the grains in THEMIS 2 are spheroids, which allows polarization observations to be reproduced. The properties of the silicate grains are based on laboratory data for silicate dust analogs measured at low temperatures \citep{Demyk2017}. THEMIS 2 is a versatile model that can combine carbon and silicate dust species with various properties (e.g., hydrogenation for carbon dust and stoichiometry for silicate dust), which naturally explains the variety of $T$ and $\beta$ values observed in the diffuse ISM \citep{Planck2013}. 

In denser regions, grains are expected to evolve. Within the THEMIS model framework, the two large grain populations first acquire a second carbon mantle (core-mantle-mantle structure, CMM), then coagulate to form aggregates covered by an icy mantle. However, the optical properties of aggregates formed from THEMIS 2 dust are not yet available. We therefore used the properties of large, spherical, and porous icy grains built from THEMIS 2 dust to mimic aggregates \cite[called ``evolved grains'' hereafter, see][for the details of the method used to calculate the optical properties]{Juvela2025}. The exact porosity and ice content influence the shape of the dust extinction curve, especially the dust optical depth ratio between FIR and \textit{J}-band ($\tau_\mathrm{FIR}/\tau_\mathrm{J}$) that influences the modeled dust NIR extinction and FIR emission \citep{Juvela2015a,Juvela2020, Juvela2025}. 
We opted for 50\% porosity and 50\% ice content (MCPI), which results in the highest $\tau_\mathrm{FIR}/\tau_\mathrm{J}$, closer to typical molecular cloud value \citep[][see also Sect.~\ref{discussion: A(J)}]{Juvela2015a, Juvela2025}.

\subsection{3D Radiative transfer modeling}\label{sec:method_SOC}

\begin{table}
\caption{Input parameters for radiative transfer models.}

\label{table:SOC_grid}  
\centering

\begin{tabular}{c|c}  
\hline       
$r_0$ [pc] &   0.04 \; 0.06 \; 0.08 \\
$\eta$  &1.0 \; 1.5\; 2.0 \\
$A_\mathrm{V}$ [mag] & 0.5 \; 1.0 \; 1.5 \\
$n_0$ [H$_2$/cm$^{3}$] & 750 \; 1500 \; 2250 \\
$a_0$ [nm] & 20 \; 40 \; 60 \\

\hline

\end{tabular}
\end{table}

L1506C was modeled using the 3D Monte Carlo radiative transfer program SOC \citep[][]{Juvela2019} that calculates the dust emission and extinction. The \textit{Herschel} and NIKA2 maps were given as observational inputs, and other user-defined input parameters described the cloud LoS structure, the radiation field and the dust models. As SOC does not require the input observation maps to have the same resolution, they were used at their nominal resolutions to achieve the best accuracy, except the 160~$\mu$m map. It was smoothed to a resolution of $18\arcsec$ in order to be consistent with the pixel size of 6$\arcsec$ for SOC modeling. We chose this pixel size as a compromise between computational time and model accuracy.

We constrained the cloud structure by adjusting $N_\mathrm{H_2}$ to fit the 350~$\mu$m intensity ($I_{350}$) map pixel by pixel. The radiation field strength $G_0$ was constrained by the ratio between the 250~$\mu$m and 500~$\mu$m intensity maps ($I_{250}/I_{500}$) averaged over the area with $I_{350}>5$~MJy/sr. With this approach, the maps of the three \textit{Herschel} bands are well fitted. We then investigated whether the NIKA2 maps are accurately predicted or not. SOC produces maps of $N_\mathrm{H_2}$, intensity and optical depth at \textit{J}-band ($\tau_\mathrm{J}$) as outputs. We converted the $\tau_\mathrm{J}$ map to extinction map with $A_\mathrm{J}=-2.5\log_{10}(e^{-\tau_\mathrm{J}})$ to compare with observations. We plotted the modeled $A_\mathrm{J}$ against observations for each 2MASS star within the FOV, and calculated the slope $k_J$ by fitting a linear function. We chose to make the comparison over the map excluding the stars inside the two core elliptical regions, because there is a clear drop of 2MASS star density toward the center, which might result in inaccuracy of the NICER estimates. The model volume is a cube with 328$^3$ cells with a side length of 6$\arcsec$. The model has a nominal resolution of 25$\arcsec$, consistent with the 350~$\mu$m map.

The cloud LoS structure was modeled with a Plummer-like function \citep{Arzoumanian2011}:
\begin{equation}
n_\mathrm{H_2}(r) = \frac{n_c}{ ( 1 +  (\frac{r}{r_0})^2  ) ^ {\eta/2}} \; ,
\label{Eq:SOC_n}
\end{equation}
where $r_0$ and $\eta$ are user-defined input parameters. The modeled filament has an asymmetric structure of a flattened cylinder (Fig.~\ref{fig:SOC_nH}). We set $n=n_\mathrm{H_2}(r)$ with distance $r$ to the plane where the filament lies (called the filament plane hereafter). In this way, we do not presume any density structure on the filament plane, which are fitted later. The central density $n_c$ is set to a constant value in the initial setting. Due to the fitting process, the final cloud profile is different from the initial one. We characterize the final cloud profiles parallel and perpendicular to the filament plane with Plummer profiles with parameters $\eta_\parallel$, $r_{0, \parallel}$ and $\eta_\perp$, $r_{0, \perp}$, respectively (Appendix~\ref{app:SOC_n}).

The external radiation field was assumed to be an isotropic interstellar radiation field \citep[ISRF,][]{Mathis1983}. Although the 160~$\mu$m map indicates some evidence of an anisotropic ISRF, we decided not to include anisotropy in the modeling, because it cannot be well constrained by our data. We assumed that L1506C is surrounded by an extended cloud outside the modeled box so that the radiation field is attenuated by an outer layer of matter characterized by an $A_\mathrm{V}$ value.

The cloud was modeled either with a single diffuse ISM type dust or with two dust populations: diffuse ISM type dust at low density and evolved grains at high density. The transition between the two types of dust was modeled with fractional abundances:
\begin{equation}
    \chi_\mathrm{diff} = \frac{1}{2} - \frac{1}{2}\tanh(4 \times (\log_{10}n_\mathrm{H_2} - \log_{10}n_0))
\end{equation}
for diffuse ISM type dust and
\begin{equation}
    \chi_\mathrm{evo} = 1-\chi_\mathrm{diff}
\end{equation}
for the evolved grains, where $n_\mathrm{H_2}$ is the modeled cloud H$_2$ number density and $n_0$ is the user-defined density threshold for the transition between the two dust components.

To study the grain growth, we varied $a_0$, the peak position of the lognormal size distribution, of the evolved grains. With a default value of $60$~nm and its value being always smaller than the minimum size, the dust size distribution has a proper hyperbolic shape (Fig.~\ref{fig:dust_size_dsitribution}). The adopted evolved grains have a fixed minimum size of $a_\mathrm{min} \sim$ 0.1~$\mu$m and maximum size of $a_\mathrm{max} = 0.7\;\mu$m. Varying $a_0$ thus changes the fraction between dust with sizes around $a_\mathrm{min}$ and $a_\mathrm{max}$. The smaller $a_0$ is, the more evolved grains have sizes around their minimum size. For THEMIS 2 diffuse type dust, we adopted its default size distribution.

To find the parameter set that best reproduces the observations, the cloud was modeled within a parameter space of all user-defined parameters mentioned before : $r_0$ and $\eta$ for the LoS density structure, $A_\mathrm{V}$ for the radiation field and $n_0$, $a_0$ for the dust models. The values of the parameters are listed in Table~\ref{table:SOC_grid}. The ranges of parameters other than $a_0$ are guided by previous studies of nearby filaments using both observational and modeling methods \citep{Arzoumanian2011, Ysard2013, Juvela2024}. We did not adopt a finer grid, because the parameters are degenerate and increasing the parameter sampling only leads to minor differences of the modeling results. Based on the fact that increasing $\eta$ has similar effect on the results as decreasing $r_0$ or increasing $n_0$, and that $A_\mathrm{V}$ is poorly constrained, expanding the parameter space further would mainly enhance the degeneracy rather than improve the fit quality (see Table~\ref{table:results_SOC_residuals} and Sect.~\ref{sec:result_SOC}). The range of $a_0$ was determined based on the fact that the modeled extinction agrees better with observations with a smaller $a_0$ (see Sect.~\ref{sec:result_SOC}). A lower limit is found at $a_0\approx20$~nm, as smaller values lead to a too small albedo compared to molecular clouds where coreshine effect is detected, as it is the case for L1506C \citep[][see also discussion in Sections ~\ref{sec:result_SOC} and \ref{sec:dust_evolution}]{Andersen2013, Steinacker2014}.

\section{Results}\label{sec:result}

In this section, we present results of the MBB fit applied either to the SEDs of each pixel on the map (Sect.~\ref{sec:results_MBB_map}) or of different substructures of L1506C (Sect.~\ref{sec:result_core_properties}). The results of 3D radiative transfer modeling are presented in Sect.~\ref{sec:result_SOC}.

\subsection{$T$, $\beta_\mathrm{all}$, $\beta_\mathrm{mm}$, and $N_\mathrm{H_2}$ maps from MBB modeling}\label{sec:results_MBB_map}

\begin{figure*}[htbp]
    \centering
    \includegraphics[width=17cm]{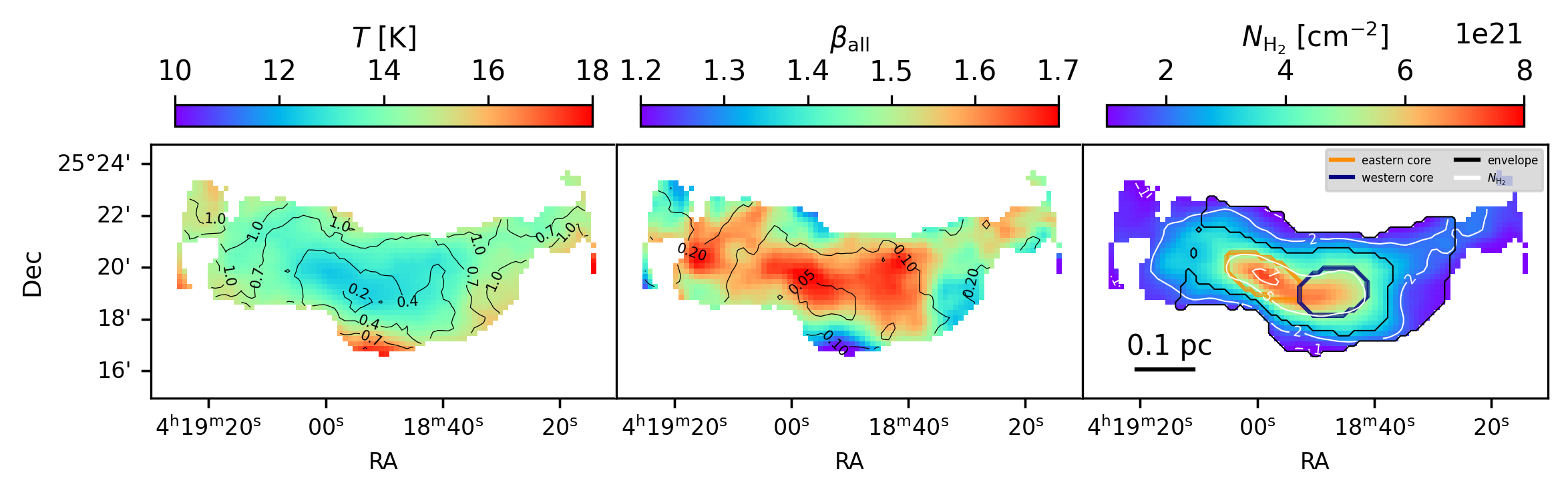}

    \caption{Temperature map (left), $\beta_\mathrm{all}$ map (middle), and $N_\mathrm{H_2}$ map (right) calculated using MBB fits to the SEDs pixel-by-pixel built using the maps at all wavelengths (160~$\mu$m to 2~mm). The black contours in the left and middle panels show the uncertainty on the $T$ and $\beta$, respectively (at levels of 0.2, 0.4, 0.7 and 1.0~K for $T$ and 0.05, 0.1 and 0.2 for $\beta$). The white contours in the right panel show $N_\mathrm{H_2}$ of levels [1, 2, 5, 7] $\times 10^{21}$~cm$^{-2}$. The region enclosed by the black contour indicates the envelope, and the two ellipses show the cores.} 
    \label{fig:MBB_fit_map}
\end{figure*}

\begin{figure}[htbp]
    \centering
    \includegraphics[width=\hsize]{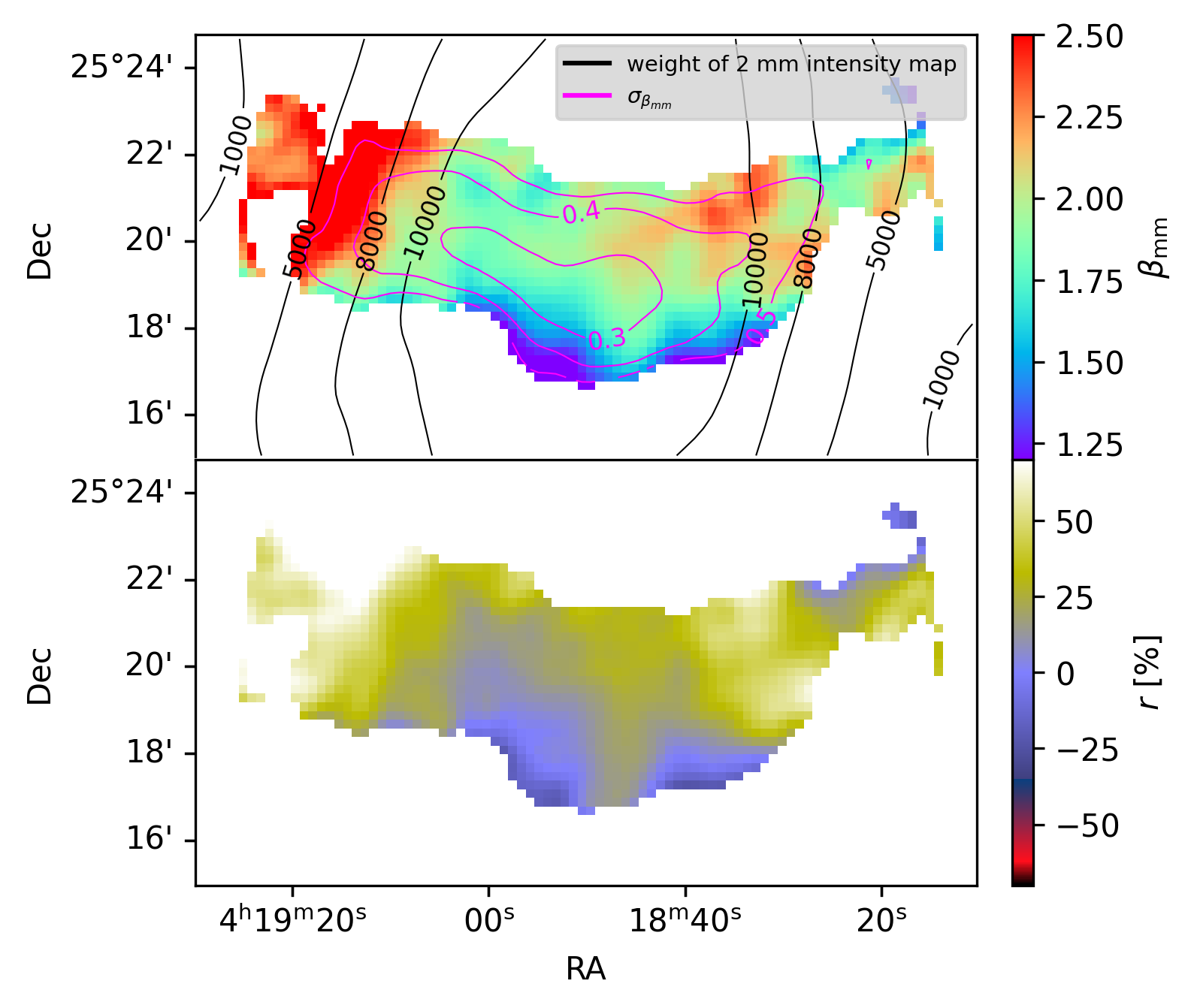}

    \caption{Top panel: Map of $\beta_\mathrm{mm}$ calculated with Eq.~\ref{equation:beta_mm}. The black contours show the weight of the 2~mm intensity map of levels [1, 5, 8, 10]~$\times 10^{3}$. The magenta contours show the uncertainty of levels 0.3, 0.4 and 0.5. Bottom panel: Percentage residual of $\beta_\mathrm{mm}$ relative to $\beta_\mathrm{all}$ ($r=$($\beta_\mathrm{mm} - \beta_\mathrm{all}$)/$\beta_\mathrm{all}$).}
    \label{fig:beta_mm}
\end{figure}

This subsection presents the results from the MBB fit with $\chi^2$ minimization. The comparison with the results obtained with hierarchical Bayesian fits is presented in Appendix \ref{app:hierarchical_bayesian}. We confirm that the results are robust with differences between the two methods below $ 10\%$ for $T$ and $\beta$, and below $15\%$ for $N_\mathrm{H_2}$.

The color temperature $T$ map (Fig.~\ref{fig:MBB_fit_map}) shows a decrease of temperature from $\sim$ 15 $-$ 16~K in the outskirts to $\sim$ 12~K in the two cores. The corresponding $\beta_\mathrm{all}$ map has a similar but reversed structure: $\beta_\mathrm{all}$ is as low as $\sim1.5$ on the edge and gradually increases to 1.7 in the center. Compared to $\beta_\mathrm{all}$, $\beta_\mathrm{mm}$ (Fig.~\ref{fig:beta_mm}) has a different structure and is larger for most of the pixels\footnote{At the eastern part, $\beta_\mathrm{mm}$ is larger by 60\% or more, and by about 40\% at the northwest part, with values $\ge2.3$. To investigate the reliability of these features at the edge, we overplotted the contour of the 2~mm intensity weight map on the $\beta_\mathrm{mm}$ map. The high $\beta_\mathrm{mm}$ region at the eastern part has a lower weight ($<8000$) than the central region. Therefore, data from this part of the map is less reliable and high $\beta_\mathrm{mm}$ at the eastern part might not be physical.}. Averaging $\beta$ over regions with 2~mm weight $>8000$ (see Fig.~\ref{fig:beta_mm}) gives $\beta_\mathrm{mm}=1.88\pm0.27$ (error bar is the standard deviation), which is larger than $\beta_\mathrm{all}=1.53\pm0.10$ in the same area. It could indicate a variation of the dust spectral index with the wavelength. This is discussed further in Sect.~\ref{sec:discussion_beta}.

The $N_\mathrm{H_2}$ map (Fig.~\ref{fig:MBB_fit_map}) shows an increase of almost one order of magnitude from the outer envelope toward the central region. The two cores, corresponding to the 250~$\mu$m intensity peaks (Fig.~\ref{fig:NIKA2_map}), have maximal $N_\mathrm{H_2}$ values of $7.4 \pm 0.9 \times 10^{21} \; \mathrm{cm}^{-2}$ for the eastern core and $6.8 \pm 0.8 \times 10^{21} \; \mathrm{cm}^{-2}$ for the western core. The structure that stretches to the northwest also has a slightly higher column density compared to its surrounding ($2 \times 10^{21} \; \mathrm{cm}^{-2}$ vs. $1 \times 10^{21} \; \mathrm{cm}^{-2}$).

\subsection{Core, envelope, and filament properties from MBB modeling}\label{sec:result_core_properties}

\begin{table}
\caption{Volume density $n(\mathrm{H}_2)$, column density $N_\mathrm{H_2}$, and mass of the cores defined in Sect.~\ref{sec:method_core_property}.} 

\label{table:core_properties}  
\centering

\begin{tabular}{ccccc}  
  
 & \textbf{eastern core} & \textbf{western core}\\
 \hline
 \hline
\multicolumn{3}{c}{B90}\\

\hline                    
$n(\mathrm{H}_2)$ [$\mathrm{cm}^{-3}$]& $(4.9 \pm 0.2) \times 10^4$ &  $(3.0 \pm 0.1) \times 10^4$\\  

$N_\mathrm{H_2}$ [$\mathrm{cm}^{-2}$]& $(5.7 \pm 0.2) \times 10^{21}$ &  $(4.6 \pm 0.2) \times 10^{21}$ \\

$M$ [$10^{-1}M_\odot$]& $7.5 \pm 1.1$ & $6.6 \pm 0.7$  \\

\hline  
\hline
\multicolumn{3}{c}{THEMIS 2 evolved grains $a_0=20$~nm}\\

\hline                    
$n(\mathrm{H}_2)$ [$\mathrm{cm}^{-3}$]& $(2.3 \pm 0.1) \times 10^4$ &  $(1.4 \pm 0.1) \times 10^4$\\  

$N_\mathrm{H_2}$ [$\mathrm{cm}^{-2}$]& $(2.7 \pm 0.1) \times 10^{21}$ &  $(2.1 \pm 0.1) \times 10^{21}$ \\

$M$ [$10^{-1}M_\odot$]& $3.5 \pm 0.5$ & $3.0 \pm 0.3$  \\

\end{tabular}
\tablefoot{The properties were calculated with different $\kappa$ values (B90 and THEMIS 2 evolved grains respectively). See Fig.~\ref{fig:dust_kappa} for comparison of $\kappa$ between the two dust models.}
\end{table}

\begin{figure}[htbp]
    \centering
    \includegraphics[width=\hsize]{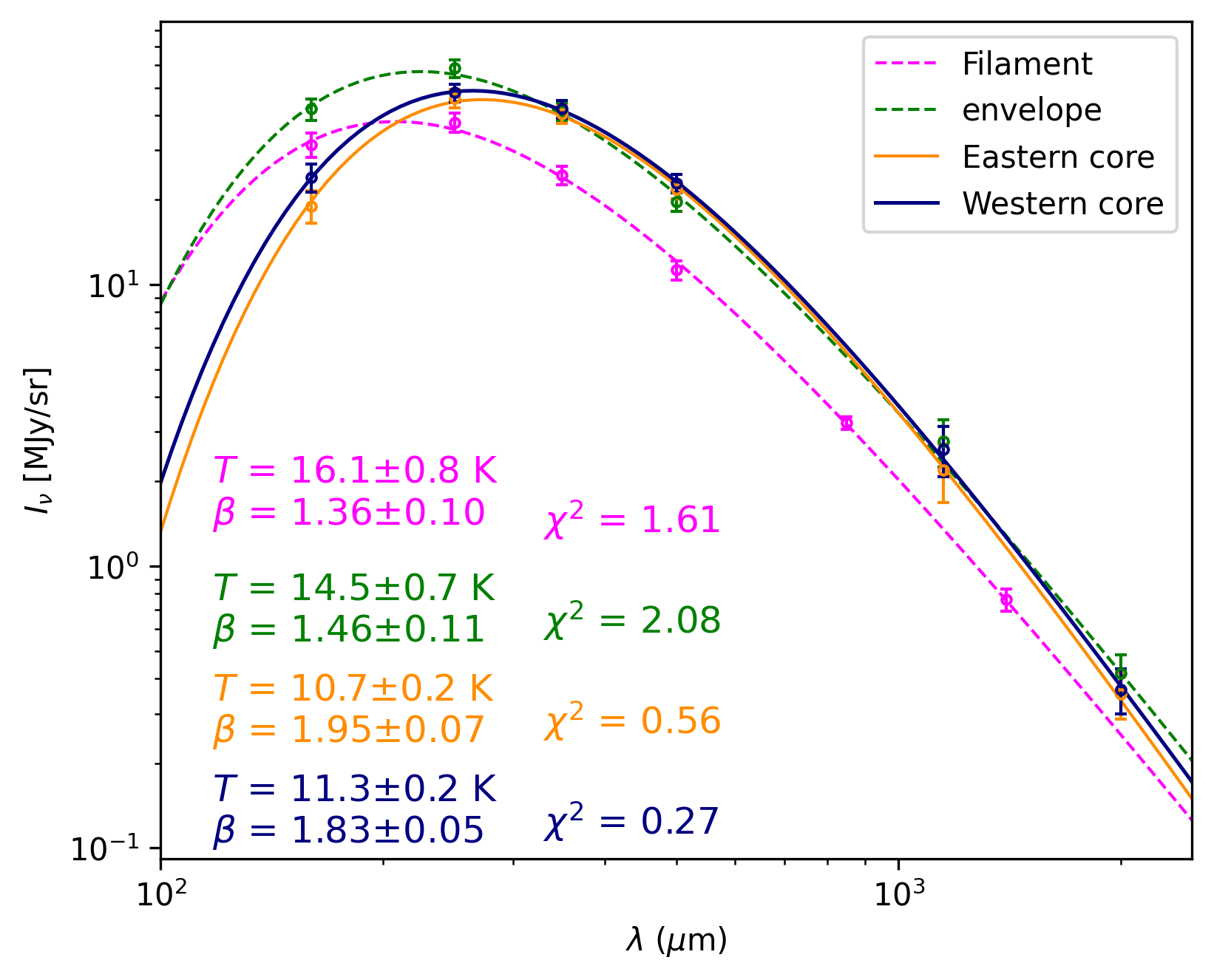}

    \caption{Modified black body fit of the SEDs of the eastern (orange) and western (blue) cores, the envelope (green dashed line), and the filament (magenta dashed line) defined in Sect.~\ref{sec:method_core_property}. The $T$, $\beta_\mathrm{all}$, and $\chi^2$ of the fits are also shown.} 
    \label{fig:core_SED}
\end{figure}

We used the $N_\mathrm{H_2}$ map to define the detailed substructures (filament, envelope and two embedded cores) of L1506C since this better traces the true core structure than the intensity maps in which the emission peaks may shift with wavelength. The cores have different sizes and can overlap slightly depending on the threshold density used to delimit them. Finally, we defined the cores by ellipses that delimit the areas where $N_\mathrm{H_2} \ge 5 \times 10^{21} \; \mathrm{cm}^{-2}$. 
The eastern and the western cores have beam deconvolved sizes of 0.13~pc $\times$ 0.05~pc and 0.11~pc $\times$ 0.07~pc respectively. The envelope is defined as the region, located inside the filament and surrounding the two cores, with $N_\mathrm{H_2} \le 3 \times 10^{21} \; \mathrm{cm}^{-2}$, S/N $\ge 4$ at each frequency, and a weight $>8000$ at 2~mm (Fig.~\ref{fig:MBB_fit_map}). Since the $N_\mathrm{H_2}$ map only shows the central part of L1506C, it is better to use the \textit{Herschel} $I_{250}$ map to define the surrounding filament, as it covers a larger area. The filament is defined as the region with $I_{250} > 55$~MJy/sr and S/N $> 12$, located outside the MBB modeling region (Fig.~\ref{fig:NIKA2_map}).
Since the area of the filament corresponds to a low S/N on the NIKA2 maps, we used instead the \textit{Planck} maps at 850~$\mu$m and 1.4~mm that were predicted at the \textit{Herschel} resolution of $36\arcsec$ using machine learning method \citep[see ][and Sect.~\ref{sec:data_reduction}]{Paradis2024}.

The SEDs of these four different substructures are shown in Fig.~\ref{fig:core_SED}. The SED intensities and the uncertainties at each wavelength are listed in Table~\ref{table:SED_values}. The eastern and western cores have similar dust properties ($T\approx11.0$~K, $\beta\approx1.90$). The filament, on the other hand, has a higher temperature of $T=16.1$~K and a lower value of $\beta=1.36$. The envelope has $T=14.5$~K and $\beta=1.46$, in between the cores and the filament. The values of $T$ and $\beta$ of the cores and the envelope agree with those resulting from hierarchical Bayesian fitting (see Appendix~\ref{app:hierarchical_bayesian}), showing their robustness. Thus, these results might indicate a change of dust properties and a sequence of dust evolution from the diffuse outer filament to the dense central core, as shown previously with the results from the MBB modeling of the map.

Table~\ref{table:core_properties} gives the $N_\mathrm{H_2}$, $n(\mathrm{H}_2)$ and mass values of the two cores calculated with two different dust models: B90, used to derive the $N_\mathrm{H_2}$ map (Fig.~\ref{fig:MBB_fit_map}), and THEMIS 2 evolved grain, used in the 3D radiative transfer modeling. Since $\kappa_{250}$ values of these two models are different by approximately a factor of 2 (Fig.~\ref{fig:dust_kappa}), the core properties are also different by the same factor.  
The $n(\mathrm{H}_2)$ values of the two cores are both comparable to or lower than previous estimates of the maximum density of L1506C based on N$_2$H$^+$ emission \citep[$5\times 10^4$~cm$^{-3}$,][]{Pagani2010}. They both have a low column density of $N_\mathrm{H_2} < 6\times 10^{21}$~cm$^{-2}$. The eastern and western cores have Jeans masses of 1.5$M_\odot$ and 1.7$M_\odot$ respectively, larger than their derived masses. Therefore, this indicates that the cores are not collapsing, and L1506C is still in a very early stage of evolution.

\subsection{Radiative transfer modeling results}\label{sec:result_SOC}

\begin{table*}
\caption{Results of radiative transfer modeling for different values of the input parameters.}

\label{table:results_SOC_residuals}  
\centering       

\setlength{\tabcolsep}{3pt}
\begin{tabular}{c|ccc|ccc|cc}  
     
$A_\mathrm{V}$ [mag] &1.0  & \textbf{1.0} & 1.0 & 1.0 & 1.0 & 1.0 & 1.0 & 1.0  \\
$n_0$ [H$_2$/cm$^{3}$] &750  & \textbf{1500} & 2250 & 750  & 1500 & 2250 &1500 &1500  \\
$a_0$ [nm] &20 & \textbf{20} & 20 & 20 & 20  & 20 &40 &60  \\
$r_0$ [pc] & 0.04 & \textbf{0.04} & 0.04 & 0.04  & 0.04 & 0.04 & 0.04 & 0.04 \\
$\eta$  &2.0 & \textbf{2.0} & 2.0  & 1.5 & 1.5 & 1.5 & 2.0 & 2.0 \\
\hline
\hline
$r_{0, \parallel}$ / $r_{0, \perp}$ [pc] & 0.07 / 0.04  & \textbf{0.07 / 0.04}  & 0.06 / 0.04 & 0.08 / 0.05  & 0.07 / 0.05 & 0.07 / 0.05 & 0.06 / 0.04 & 0.06 / 0.04 \\
$\eta_\parallel$ / $\eta_\perp$ & 2.0 / 3.3  & \textbf{1.9 / 3.3} & 1.8 / 3.3  & 1.9 / 2.9  & 1.7 / 2.9 & 1.7 / 2.9 & 1.9 / 3.4 & 1.9 / 3.4 \\
$n_c$ [$10^4$ H$_2$/cm$^{3}$] & 3.1   & \textbf{3.4}  & 3.7   & 1.8   & 2.1  & 2.3  & 3.7  & 3.9 \\

$G_0$ & 1.51 & \textbf{1.55} & 1.56 & 1.36 & 1.40 & 1.41 & 1.53 & 1.52\\

\hline
\hline

\shortstack{ core residual \\ 160~$\mu$m [\%]} & $-27.9$  $\pm 10.0$ &  $\mathbf{-22.0}$  $\mathbf{\pm10.1}$ &  $-19.9$  $\pm 8.7$ &$-40.6$  $\pm 10.3$  &$-31.8$  $\pm 11.2$  & $-28.1$  $\pm 9.6$  & $-23.5$  $\pm 9.3$ & $-24.7$  $\pm 9.1$   \\ [3pt]

\shortstack{1.15~mm[\%]} & 12.0  $\pm 2.67$ &  $\mathbf{9.4}$  $\mathbf{\pm3.3}$  & $-7.2$ $\pm 3.3$  & $19.2$  $\pm 2.9$  & $16.2$  $\pm 2.9$  & $13.9$  $\pm 2.8$  & $7.1$  $\pm 3.5$  &  $4.8$  $\pm 3.8$ \\[3pt]

\shortstack{2~mm [\%]} & $-0.2$  $\pm 6.5$ &  $\mathbf{-4.0}$  $\mathbf{\pm6.6}$  & $-6.8$  $\pm 7.0$  & $9.5$  $\pm 5.0$  & $4.9$  $\pm 6.3$  & $1.5$  $\pm 6.3$ & $-7.5$  $\pm 7.0$  &  $-10.3$ $\pm 7.3$  \\ [3pt]

\hline
\hline

$k_{J}$  &  2.66 $\pm$ 0.03   &  \textbf{2.77}  $\mathbf{\pm 0.03}$  & 2.85 $\pm$ 0.03  & 2.37  $\pm$ 0.02   & 2.52  $\pm$ 0.03  &  2.61 $\pm$ 0.03 &  3.93 $\pm$ 0.04 &  4.80  $\pm$ 0.04   \\ [3pt]

max($I_\mathrm{sca}$) [MJy/sr] & 0.026  & \textbf{0.027}  & 0.026 & 0.028 & 0.031 & 0.032  &  0.055 & 0.086 \\

\end{tabular}
\tablefoot{The first five rows show the input parameters. The following four rows show the maximum H$_2$ density $n_c$, $\eta_\parallel$, $r_{0, \parallel}$, $\eta_\perp$ and $r_{0, \perp}$ of the final modeled cloud and $G_0$. The next four rows show the residuals in percentage obtained in the cores areas ((observation $-$ model)/observation): they are estimated using the peak position and standard deviation of the Gaussian distribution fitted to the histogram of intensity residuals at different wavelengths (example shown on the bottom row of Fig.~\ref{fig:SOC_modeling_I}). The last two rows show the correlation slope ($k_J$) derived when plotting the modeled $A_\mathrm{J}$ against WIRCam $A_\mathrm{J}$ for each 2MASS star (see Fig.~\ref{fig:SOC_modeling_AJ} and related discussion), and the maximum scattered intensity within the \textit{Spitzer} IRAC1 3.6~$\mu$m filter (Fig.~\ref{fig:SOC_scattering}). The parameter set of the model discussed in the text is shown in bold.}
\end{table*}

\begin{figure*}[htbp]

    \centering
    \includegraphics[width=18cm]{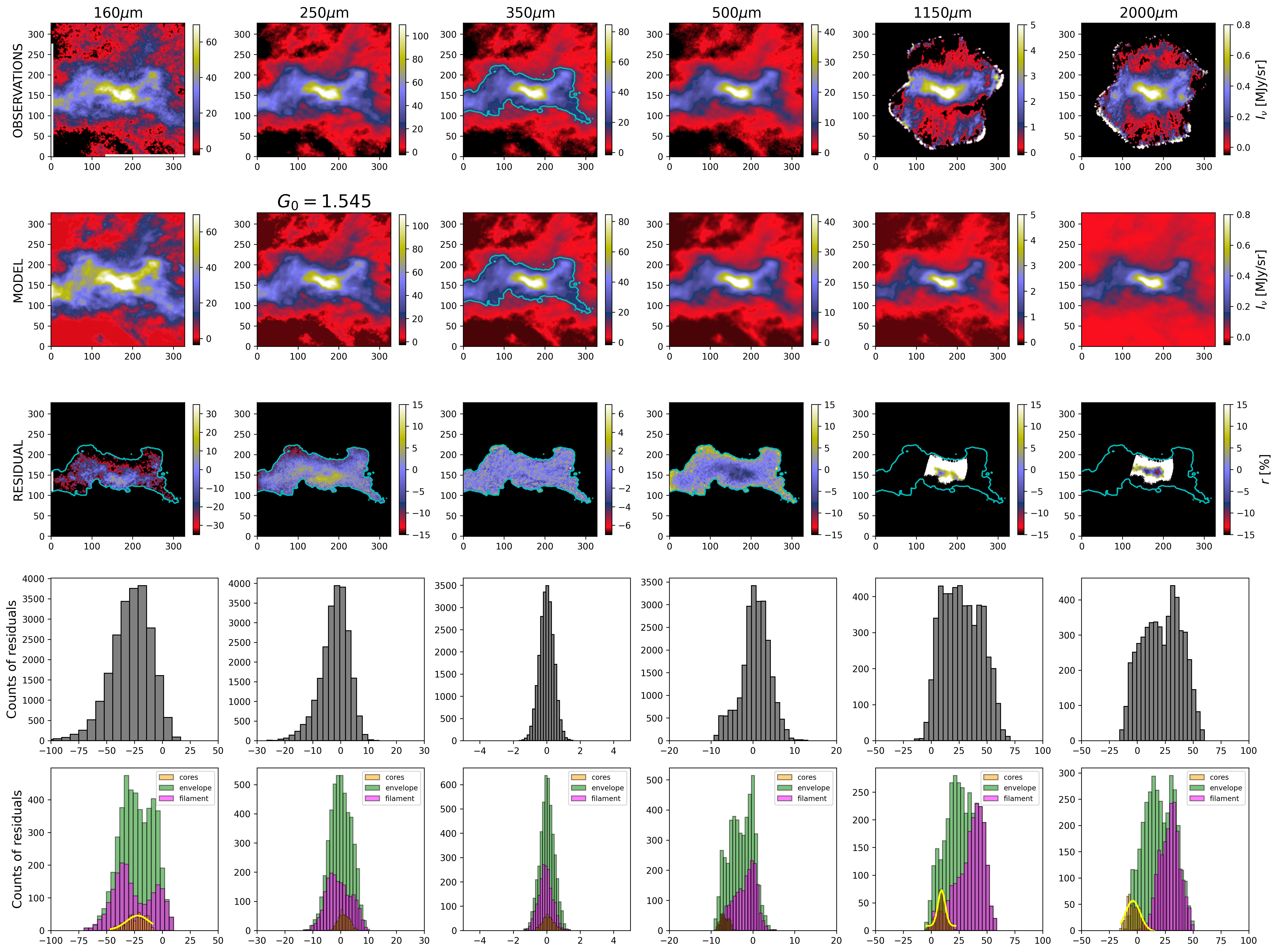}
    \caption{Modeling results of intensity maps with the chosen parameters (see Table~\ref{table:results_SOC_residuals}). The first row shows observed maps at each wavelengths, the second row shows the modeled maps, the third row shows the relative errors (called ``residuals'') of intensity in percentage ((observation $-$ model)/observation), so positive error means underestimation), the fourth row shows the histogram of the residuals displayed on the third line, and the last row shows the histogram of residuals within different regions defined in Sect.~\ref{sec:result_core_properties} (i.e., orange for cores, green for the envelope and magenta for the filament). Residuals are calculated within the cyan contour ($I_{350\mu m}>15$~MJy/sr) for the {\it Herschel} bands and within regions with $I_{1mm}>1.5$~MJy/sr and 2~mm weight $>$~8000 for the NIKA2 bands. We fitted Gaussian functions to the histograms of the residuals in the core region for 160~$\mu$m, 1.15~mm and 2~mm, which are shown in yellow on the last row. The fitted peak position and standard deviation are listed in Table~\ref{table:results_SOC_residuals}. The fitted $G_0$ is shown between the first and second row.} 
    \label{fig:SOC_modeling_I}
\end{figure*}

\begin{figure}[t]
    \centering
    \includegraphics[width=\hsize]{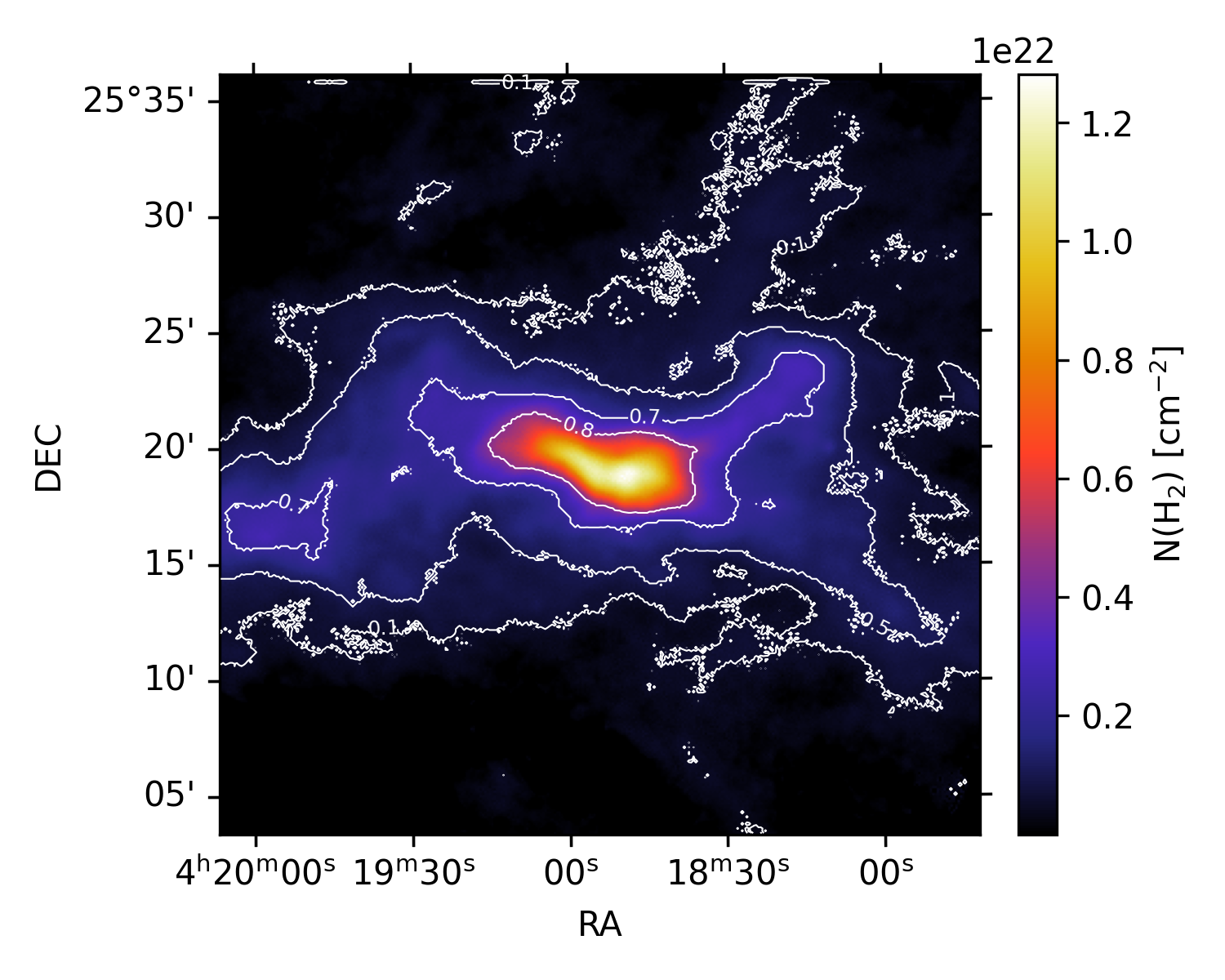}
    \includegraphics[width=\hsize]{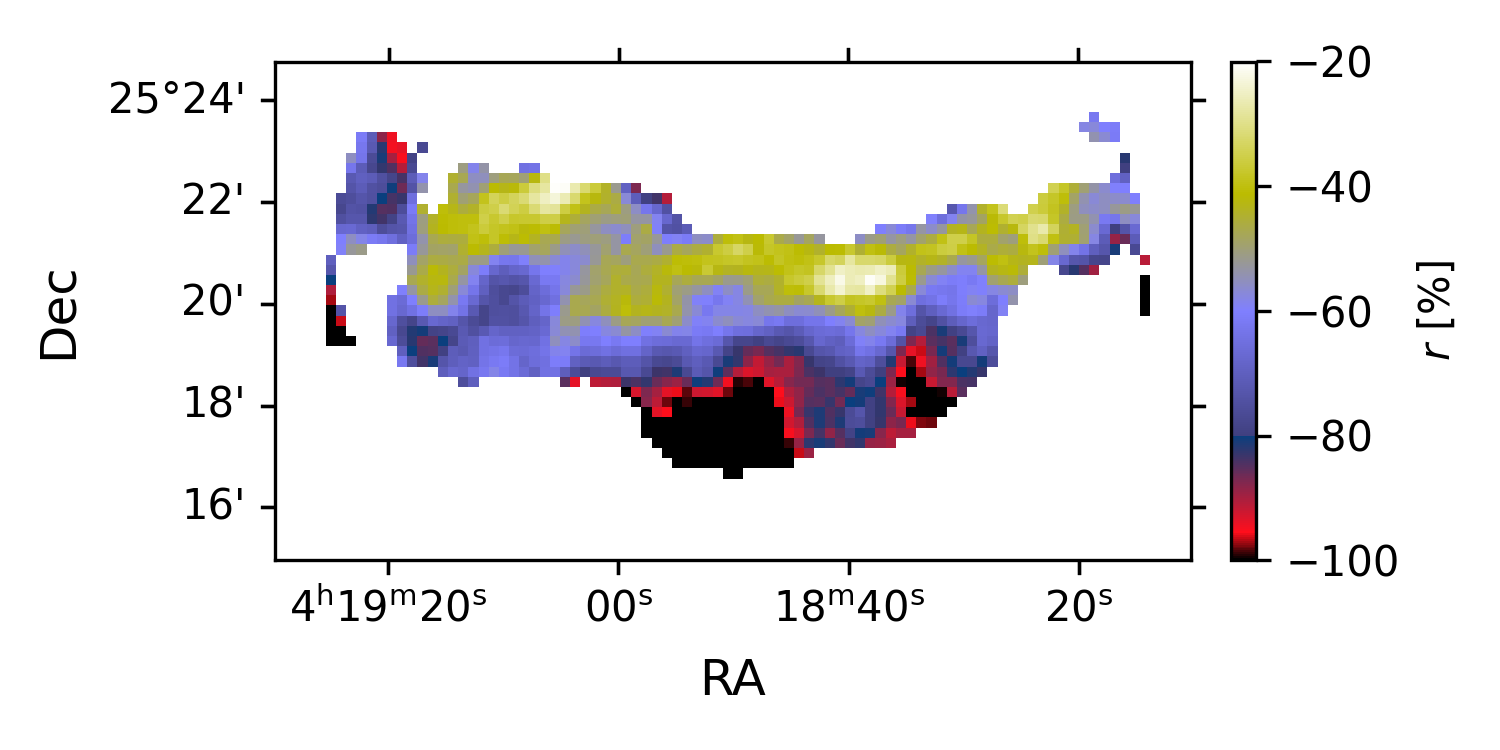}

    \caption{\textit{Top panel:} Map of $N_\mathrm{H_2}$ derived using the radiative transfer model for the chosen parameter set (see Table~\ref{table:results_SOC_residuals}). The white contours show the fraction of dust along the LoS that is in evolved grains (at levels of 0.1, 0.5, 0.7, and 0.8). \textit{Bottom panel:} Residual of the $N_\mathrm{H_2}$ modeled with SOC compared to the one from MBB modeling (($N_\mathrm{H_2}$(MBB) $-$ $N_\mathrm{H_2}$(SOC))/$N_\mathrm{H_2}$(MBB)).}
    \label{fig:SOC_modeling_NH}
\end{figure}

\begin{figure*}[htbp]
    \centering
    \includegraphics[width=0.72\linewidth]{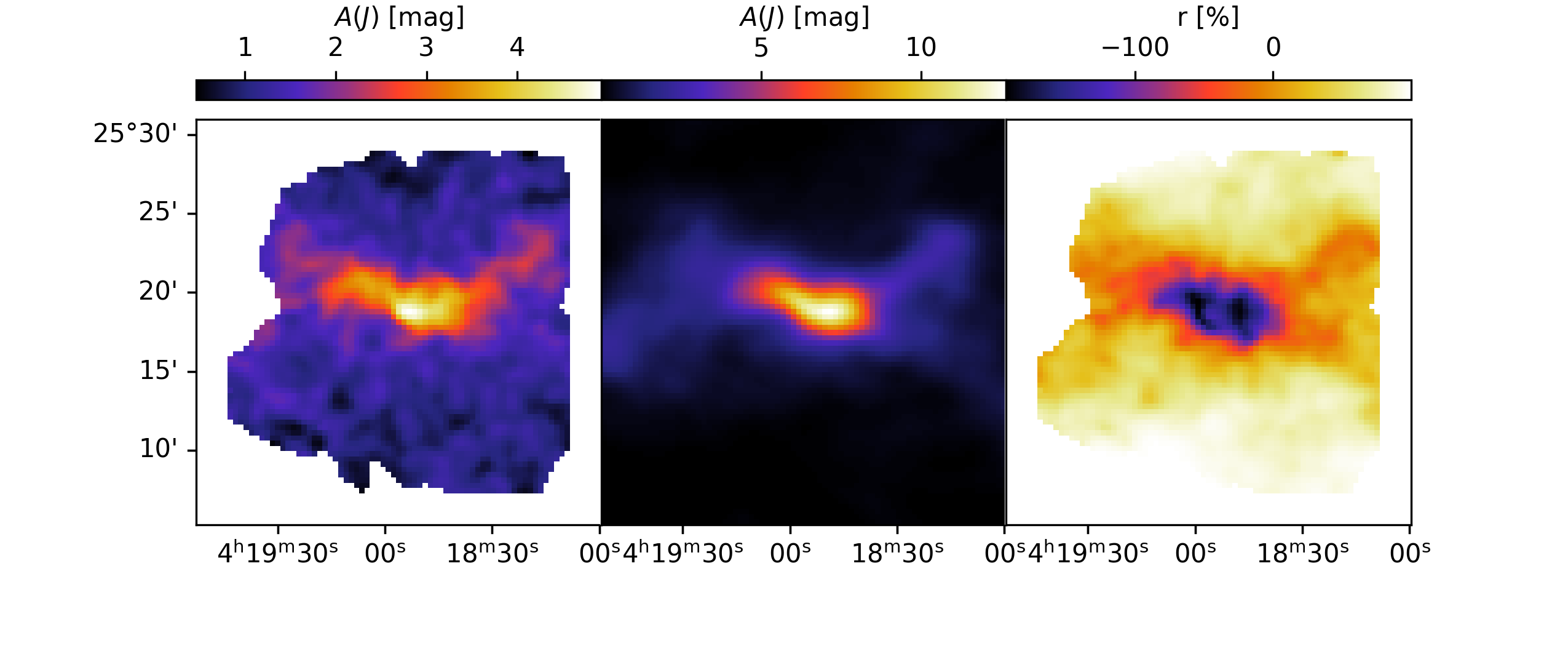}
    \hspace{-10mm}
    \raisebox{2mm}{
    \includegraphics[width=0.32\linewidth]{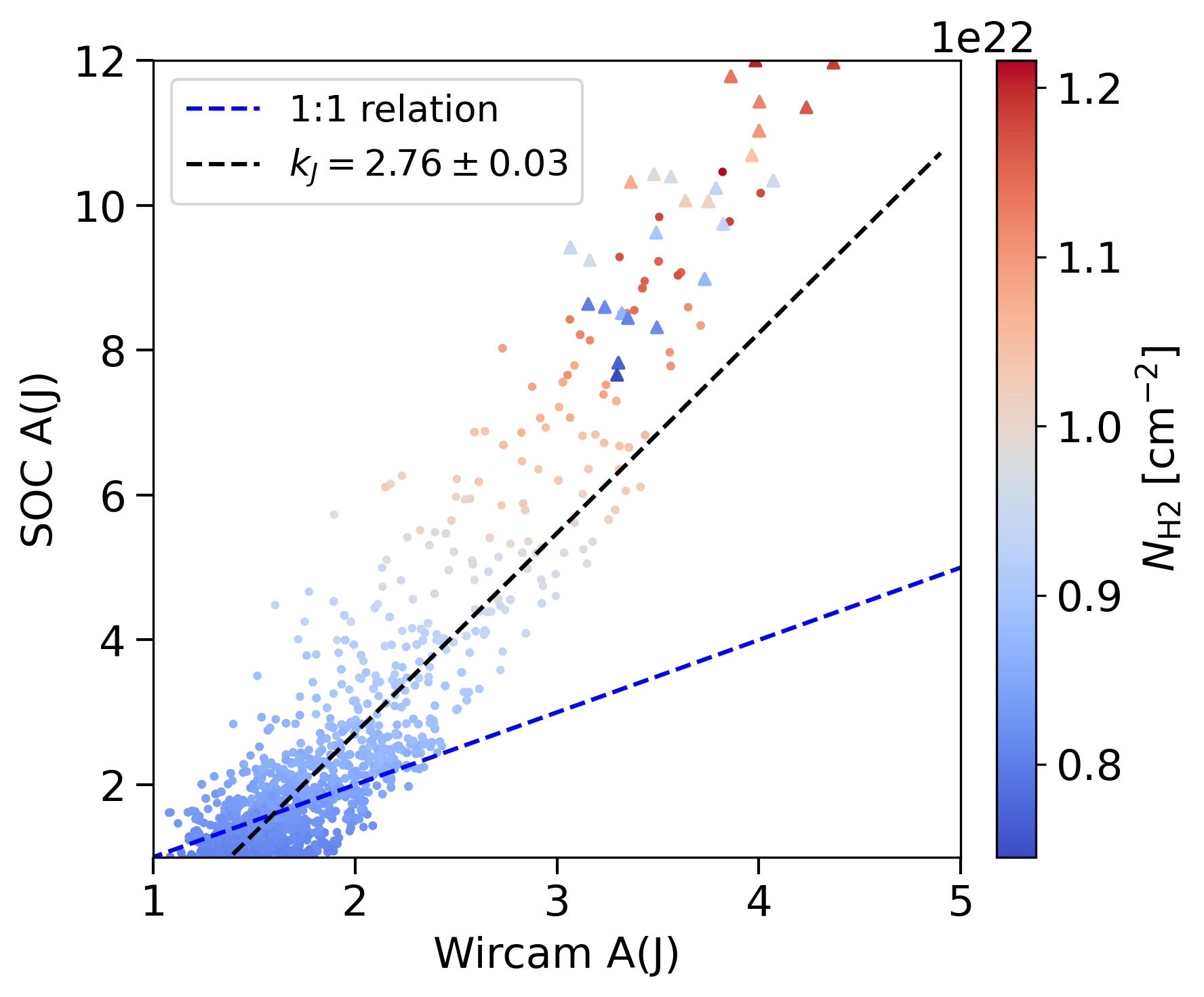}}

    \caption{Comparison of the extinction modeled with SOC and observed with WIRCam. From left to right, the panels are as follows. {\it First panel}: Map of $A_\mathrm{J}$ calculated from WIRCam observations (see Sect.~\ref{sec:wircam}).  
    {\it Second panel}: Map of $A_\mathrm{J}$ modeled with SOC and the chosen parameter set (see Table~\ref{table:results_SOC_residuals}), smoothed to the same resolution as the WIRCam extinction map (1$\arcmin$). {\it Third panel}: Residual ((observation $-$ model) / observation). {\it Fourth panel}: $A_\mathrm{J}$ modeled by SOC plotted against $A_\mathrm{J}$ from WIRCam observations for every 2MASS star used to estimate WIRCam $A_\mathrm{J}$. The colorbar is the modeled $N_\mathrm{H_2}$ along the LoS of the 2MASS stars. The slope $k_J$ is shown on the top left. Triangles show the stars that are within the two central core ellipses and excluded when calculating $k_J$.} 
    \label{fig:SOC_modeling_AJ}
\end{figure*}

\begin{figure}[htbp]
    \centering
    \includegraphics[width=\hsize]{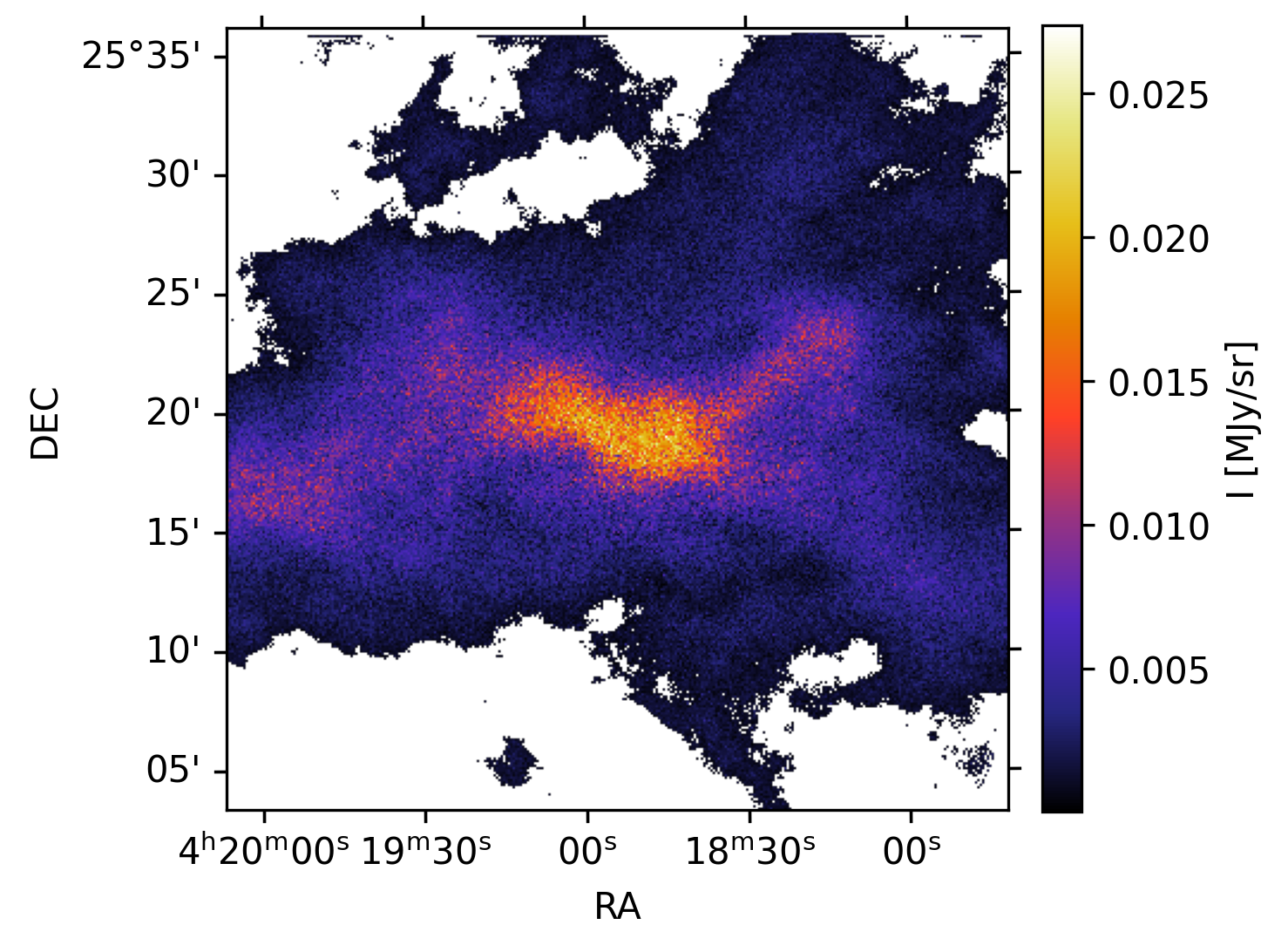}

    \caption{Map of the scattered light within the \textit{Spitzer} IRAC1 3.6~$\mu$m filter modeled by SOC with the chosen parameter set. The coreshine of L1506C observed by \textit{Spitzer} IRAC1 is $0.04\pm0.03$~MJy/sr above the background at the peak position \citep{Steinacker2014}.} 
    \label{fig:SOC_scattering}
\end{figure}

To find the best model reproducing the observations of L1506C, we performed radiative transfer modeling for various combinations of the input parameters presented in Table~\ref{table:SOC_grid}. The results are presented in Table~\ref{table:results_SOC_residuals} only for certain combinations with two dust models. They show that the residuals between the observations and the model for the \textit{Herschel}, NIKA2, and to a lesser extend for the WIRCam observations ($k_J$, as defined in Sect.~\ref{sec:method_SOC}) have similar values for a wide range of parameter values, illustrating that some parameters are degenerate and cannot be well constrained.
 
The least constrained parameter is $A_\mathrm{V}$, the extinction of the outer layer that attenuates the ISRF. When $A_\mathrm{V}$ increases, the shielding of the clump increases. Consequently, to reproduce the observed emission, $G_0$ increases and $N_\mathrm{H_2}$ and $A_\mathrm{J}$ decrease. However, the ISRF being independent of the cloud emission, it is difficult to determine $A_\mathrm{V}$ using intensity and extinction maps. \cite{Juvela2024} modeled the whole filament LDN~1506 using SOC with \textit{Herschel} and 2MASS data and argued that $A_\mathrm{V} > 1$ would imply a too high $G_0$ for Taurus. The author also found $A_\mathrm{V} \approx 1$ around the modeled area based on \textit{Planck} optical depth maps. Therefore, we adopted $A_\mathrm{V} = 1$.

The best constrained parameter is $a_0$, the peak of the grain size distribution. This parameter has a strong influence on the extinction but not on the emission. Indeed, for micron-size grains, the opacity in the FIR and millimeter ranges does not depend on $a_0$ whereas, in the NIR range, the grain size being similar to the wavelength, scattering can no longer be neglected, which leads to a variation of $\kappa$ with $a_0$ (Fig.~\ref{fig:dust_kappa}). For $a_0=20$~nm, the modeled \textit{J}-band extinction $A_\mathrm{J}$ is still overestimated by a factor $k_J \sim2.4-2.8$, independently of the other parameter values. The factor reaches almost 5 when $a_0$ is increased to 60~nm. Although $k_J$ can be further lowered with a smaller $a_0$, we adopted 20~nm as a lower limit as explained in Sect.~\ref{sec:method_SOC}.

The slope, $\eta$, and the flat central radius, $r_0$, of the Plummer LoS profile have a rather pronounced influence on the modeling results. As increasing $r_0$ and decreasing $\eta$ both smooth the cloud geometry in similar ways, these two parameters are degenerate and difficult to constrain individually. When $\eta$ increases or $r_0$ decreases, the modeled intensity, extinction and central volume density all increase. A possible reason for this is that, due to the steeper density profile, $n_\mathrm{H_2}$ drops quickly toward the edge; consequently a higher density and $G_0$ are needed to emit the same $I_{350}$, which raises the overall extinction and intensity at other wavelengths. 

The influence of the threshold density for the transition between diffuse type dust and evolved grains, $n_0$, has a direct impact on the total dust emission. The fraction of evolved grains decreases with higher $n_0$. Therefore, since evolved grains are more emissive than diffuse grains, the total emission decreases and a higher density ($n_c$) is required to compensate and fit the emission. The effect is similar to increasing $\eta$ and decreasing $r_0$. The density of the cores can be used to constrain $\eta$, $r_0$ and $n_0$, since $n_c$, the maximum density in a modeled cell, should be consistent with MBB modeling results and cannot be less than the average density in the core region, $n_\mathrm{H_2} = 2.3 \times 10^4$ cm$^{-3}$ for the densest core (Table~\ref{table:core_properties}).

We present below the results obtained with the parameter set in bold in Table~\ref{table:results_SOC_residuals} but point out that other parameter sets, particularly those differing only in their $n_0$ values (first and third columns from the left in Table~\ref{table:results_SOC_residuals}), also reproduce the observations reasonably well. This choice is a compromise between the various criteria on the density ($n_c \approx 2.3 - 5\times 10^4\;\mathrm{cm}^{-3}$) and the residuals (minimizing $k_J$ and the \textit{Herschel} and NIKA2 residuals). The chosen parameters correspond to $A_\mathrm{V} = 1$ and $a_0=20$~nm. For $\eta$, $r_0$ and $n_0$, which are degenerate, we opted for $\eta=2$, $r_0=0.04$ and $n_0=1500$~cm$^{-3}$ with $\eta_\parallel=1.9$, $r_{0, \parallel}=0.07$~pc, $\eta_\perp=3.3$ and $r_{0, \perp}=0.04$~pc, so that the fraction of evolved grain is not too large in the filament (see below). The asymmetric structure of the cloud (Fig.~\ref{fig:SOC_nH}) is a consequence of the chosen initial setting (Eq.~\ref{Eq:SOC_n}). If we model the cloud with an initial cylindrical geometry, the final geometry changes to $r_{0, \parallel}=r_{0, \perp}=0.05$~pc and $\eta_\parallel=\eta_\perp=2.7$. However, the modeled emission and extinction change by $<$10\%, showing that the cloud geometry is highly degenerate and cannot be constrained precisely.

Figure~\ref{fig:SOC_modeling_I} shows the modeled intensity and residuals compared to the observations for the chosen parameter set. We plotted the overall histograms of the residuals at each wavelength and the histograms of the different regions (core, envelope and filament) as defined in Sect.~\ref{sec:result_core_properties}. The overall histograms have the smallest spread and are centered around zero for $I_{250}$, $I_{350}$ and $I_{500}$ that were used to fit $N_\mathrm{H_2}$ and $G_0$. While the overall histogram at 160~$\mu$m mostly has negative values (overestimation of the flux), the ones at millimeter wavelengths mostly have positive values (underestimation of the flux) for all parameter sets in Table~\ref{table:results_SOC_residuals}. This is due to the fact that $N_\mathrm{H_2}$ is fitted based on $I_{350}$, as explained in \cite{Juvela2025}. Histograms of different regions, however, have different properties. For example, the filament residual histogram at 160~$\mu$m is asymmetric, with intensity being underestimated in the southern part and overestimated in the northern part. Possible explanations could be an anisotropic radiation field with stronger radiation coming from the south, or an asymmetrical shape of the cloud that cannot be well constrained by our model assumptions. The core histograms at 160~$\mu$m and millimeter wavelengths are centered closer to zero than the overall histogram ($\lesssim 22$\% for the former and $\lesssim 10$\% for the latter), showing that the core intensities are well reproduced. In general, fitting the observed intensities at 250, 350 and 500~$\mu$m well reproduces the intensities at millimeter wavelengths in the area of the cores, but underestimates the intensities in the filament and envelope area.

The model $N_\mathrm{H_2}$ has a maximum value of $1.3 \times 10^{22}$~cm$^{-2}$ (Fig.~\ref{fig:SOC_modeling_NH}), about 1.7 times higher than the maximum $N_\mathrm{H_2}$ derived from MBB fit toward the cores, indicating the importance of radiative transfer modeling to correctly constrain $N_\mathrm{H_2}$. As $T$ derived from the MBB fit is an effective mean temperature over the LoS, which is higher than the true dust temperature of the cores, $N_\mathrm{H_2}$ from MBB fit is also smaller than the true column density toward the cores. Taking into account radiative transfer, the evolved grains have a minimum temperature of 8~K at the core center for the chosen parameter set, a value that varies between 7 and 9~K depending on the set of input parameters. In all cases, the modeled dust minimum $T$ is about 2 to 4~K colder than the MBB color temperature. It is also consistent with the typical floor temperature set by cosmic ray heating of gas and grains deep inside prestellar cores \citep[5-6~K,][]{Goldsmith2001, Hocuk2017, Ivlev2019}. The SOC modeled $N_\mathrm{H_2}$ agrees within a factor of 1.5 with the values derived from the chemical modeling of CO and N$_2$H$^+$ \citep[$2 \times 10^{22}$~cm$^{-2}$,][]{Pagani2010}. 

Figure~\ref{fig:SOC_modeling_NH} shows the fraction of evolved grains along the LoS, overplotted on the $N_\mathrm{H_2}$ map. This fraction is defined as the fraction of $N_\mathrm{H_2}$ where only evolved grains are present. With the chosen set of parameters, the central part of L1506C is dominated by evolved grains by $>80$\%. This value changes to 85\% or 75\% if $n_0$ is varied to 750 or 2250~cm$^{-3}$, respectively. Moreover, it is noteworthy that even in the filament ($N_\mathrm{H_2} \gtrsim 1\times 10^{21}$~cm$^{-2}$) already 50\% of $N_\mathrm{H_2}$ corresponds to evolved grains (70\% or 40\% with $n_0=750$ or 2250~cm$^{-3}$). The fraction of evolved grains increases rapidly between the 0.1 and 0.5 contours (Fig.~\ref{fig:SOC_modeling_NH}). This suggests that grains evolve at a very early stage of star formation where gravitational collapse has not yet started. This is discussed further in Sect.~\ref{sec:dust_evolution}.

Figure~\ref{fig:SOC_modeling_AJ} shows that within the filament area, the modeled $A_\mathrm{J}$ has a residual around 50\% compared to the WIRCam $A_\mathrm{J}$. Within the envelope area, however, the residual increases over 100\%. We plotted the modeled $A_\mathrm{J}$ against WIRCam $A_\mathrm{J}$ and fitted a slope to the data points, neglecting the ones within the cores area, which shows that $A_\mathrm{J}$ is overestimated by a factor $\sim 2.8$. The higher the $N_\mathrm{H_2}$, the more $A_\mathrm{J}$ is overestimated. This could be partly due to the lower 2MASS star density at high density regions, implying that the WIRCam $A_\mathrm{J}$ is less accurate and underestimated in the center. However, the dust properties (specific dust model used, size distribution, etc.) have a strong impact on the extinction and thus on the $A_\mathrm{J}$ overestimation. It is discussed in detail in Sect.~\ref{discussion: A(J)}.

L1506C was also observed as part of the Taurus \textit{Spitzer} survey \citep[PI: P.L. Padgett,][]{Rebull2010} and a signature of coreshine with a peak brightness of $0.04\pm0.03$~MJy/sr above background  measured over a $10\arcsec$ square was observed with the \textit{Spitzer} IRAC1 instrument at 3.6~$\mu$m \citep{Steinacker2014}. For comparison purposes, we modeled scattering with SOC at 10 wavelengths (0.8~$\mu$m, 2.0~$\mu$m, 2.8~$\mu$m, 3.6~$\mu$m, 4.8~$\mu$m, 5.6~$\mu$m, 7.7~$\mu$m and 10.0~$\mu$m). We fitted the resulting scattering SED for each pixel with a lognormal distribution as a function of wavelength. The modeled scattered intensity within the IRAC1 3.6~$\mu$m filter was finally calculated as the average between the fitted intensity at 3.1~$\mu$m and 3.9~$\mu$m. The resulting map of the scattered intensity is shown in Fig.~\ref{fig:SOC_scattering}. With a maximum intensity (max($I_\mathrm{sca}$) in Table~\ref{table:results_SOC_residuals}) of 0.027~MJy/sr, we can reproduce the observed level of coreshine within 1~$\sigma$ for the parameter sets of the three first columns of Table~\ref{table:results_SOC_residuals}.

\section{Discussion}\label{sec:discussion}

In this section, we discuss variations of $\beta_\mathrm{all}$ and $\beta_\mathrm{mm}$, and their correlations with $T$ (Sect.~\ref{sec:discussion_beta}). Evidence of dust evolution from observations and radiative transfer modeling is discussed in Sect.~\ref{sec:dust_evolution}, and the reason for the $A_\mathrm{J}$ overestimation in the modeling is discussed in Sect.~\ref{discussion: A(J)}.

\subsection{Dust spectral index variation}\label{sec:discussion_beta}

\begin{figure}[t]
    \centering
    \includegraphics[width=\hsize]{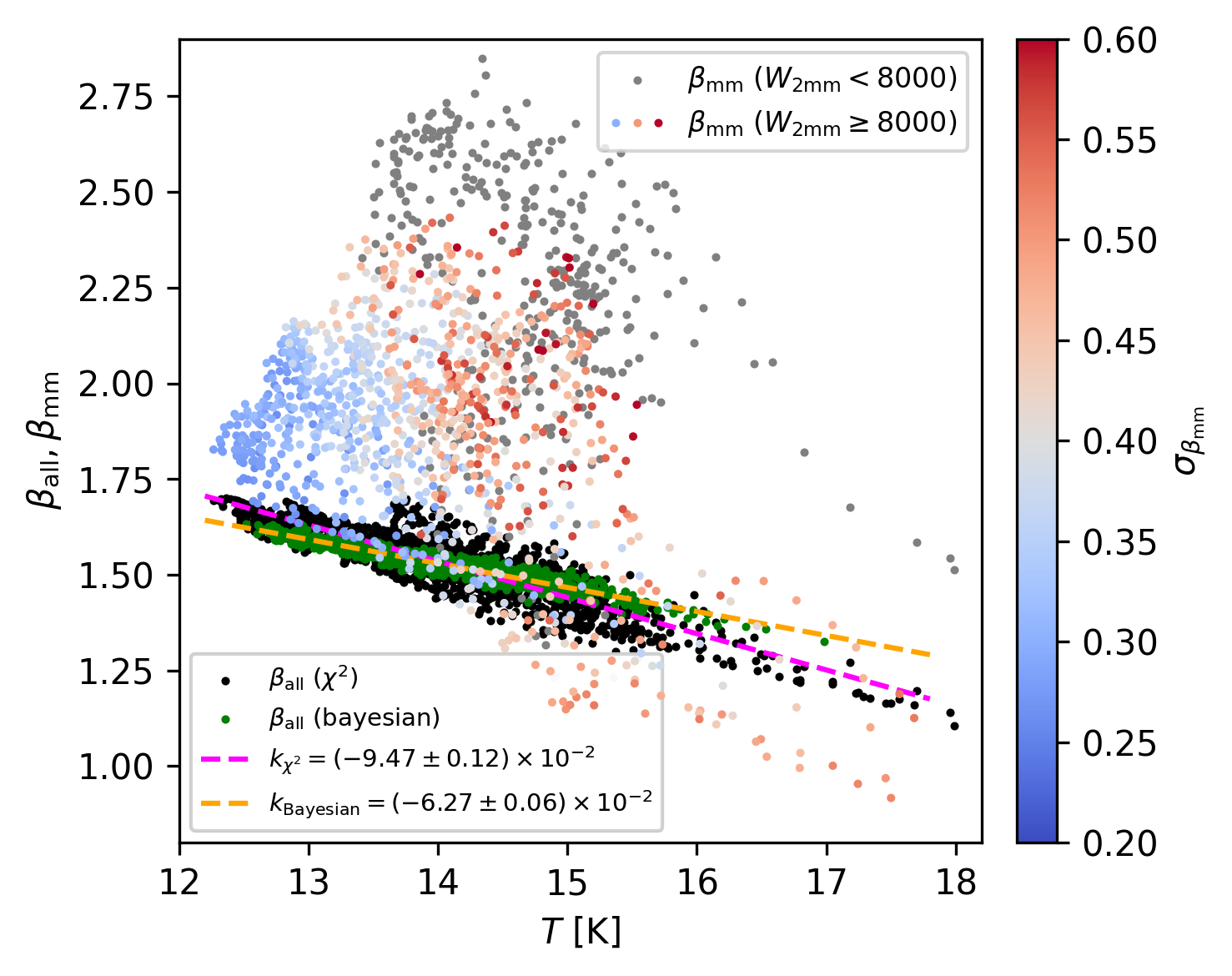}

    \caption{Spectral index as a function of temperature. For pixels with S/N $\ge4$, values of $\beta_\mathrm{all}$ derived with a $\chi^2$ minimization (black circles) and a hierarchical Bayesian method (green circles) are shown. The uncertainty of the $\beta_\mathrm{all}$ values are smaller than 0.3 for all points and for both methods. The values of the $\beta_\mathrm{mm}$ of pixels with 2 mm weight ($W_\mathrm{2mm}$) $<8000$ (gray circles) and the $\beta_\mathrm{mm}$ of pixels with $W_\mathrm{2mm} \geq 8000$ (colored circles with uncertainty indicated by the color bar) are plotted for all pixels with S/N $\ge4$. The linear least square fit of the relation $\beta(T)=kT + m$ is shown for the $\chi^2$ (magenta dashed line) and the Bayesian (orange dashed lines) method. The fitted $k$ value for both methods are shown at the bottom-left corner.} 
    \label{fig:T_beta_mm}
\end{figure}

As shown in the $T$ and $\beta$ maps (Fig.~\ref{fig:MBB_fit_map}) and from the MBB modeling of the SEDs of the cores and filament (Fig.~\ref{fig:core_SED}), $T$ and $\beta_\mathrm{all}$ show spatial variations beyond their uncertainties. $T$ decreases and $\beta_\mathrm{all}$ increases from the outer filament toward the central cores, indicating a change of grain properties, possibly due to grain evolution. This increase of $\beta_\mathrm{all}$ is consistent with dust models: evolved grains have larger $\beta$ (between 160~$\mu$m and 2~mm) than THEMIS 2 diffuse type dust.

Figure~\ref{fig:MBB_fit_map} shows that the spatial variations of $T$ and $\beta$ are anticorrelated. This anticorrelation is plotted in Fig.~\ref{fig:T_beta_mm}. We performed a linear least-square fit of the form $\beta(T)=kT + m$ to the $T-\beta$ relations derived either from the $\chi^2$ minimization method or hierarchical Bayesian method, restricting the analysis to pixels with S/N $\ge 4$. 
The anticorrelation is present for both methods and may be due to intrinsic grain properties. The fitted slope for $\chi^2$ minimization, $k_{\chi^2}=-0.09$, is steeper than the slope for hierarchical Bayesian, $k_\mathrm{Bayesian}=-0.06$, which is expected as the $\chi^2$ minimization method is known to artificially exaggerate the $T-\beta$ anticorrelation due to the effect of noise and flux uncertainty \citep{Shetty2009, Planck2011, Juvela2013}. The value of $k_\mathrm{Bayesian}$ agrees well with literature, e.g., $k \sim -0.05$ from \textit{Herschel} SPIRE and \textit{Planck} data of various galactic cold clumps \citep{Juvela2015b}.

On the other hand, in the millimeter range, $\beta_\mathrm{mm}$ shows significant scatter and there is no clear anticorrelation with $T$ (Fig.~\ref{fig:T_beta_mm}). This cannot be explained by uncertainties in the dust temperature, since computing $\beta_\mathrm{mm}$ with $T$ derived from all wavelength bands results in changes of no more than 5\%, even when $T$ changes by over 4~K. Changes in $\beta_\mathrm{mm}$ with respect to temperature and wavelength are predicted in the framework of the two-level-system (TLS) theory \citep{Meny2007} but were not observed in silicate dust analogs on such a narrow temperature range \citep{Mennella1998, Agladze1996, Demyk2017}. In L1506, $\beta_\mathrm{mm}$ appears higher than $\beta_\mathrm{all}$ for most of the pixels, indicating a change in the dust emissivity behavior with wavelengths. Such a behavior is compatible with some of the silicate grains used in the THEMIS 2 dust model \citep{Ysard2024}.

The average value of $\beta_\mathrm{mm}$ within the area with 2~mm weight $>$~8000, $1.88\pm0.27$ (error bar denotes the standard deviation), is lower than the values, obtained for other starless or prestellar cores from NIKA2 data, of $2.4\pm 0.3$ \citep{Bracco2017}, $2.01\pm0.48$ \citep{Scibelli2023}, and $2.0\pm0.3$ \citep{Nguyen-Luong2024}. Possible explanation is that we used the reconstructed maps in which the extended emission from the filament is recovered, so that $\beta_\mathrm{mm}$ is attenuated by the diffuse ISM component along the LoS which has a lower $\beta$. If we calculate $\beta_\mathrm{mm}$ within the area with 2~mm weight $>$~8000 with the original NIKA2 maps, it has a value of $2.15\pm0.39$, consistent with literature values. 
Although $\beta_\mathrm{mm}$ values do show some spatial variations (Fig.~\ref{fig:beta_mm}), the variations are within error bars (mean $\beta_\mathrm{mm}=1.86\pm 0.33$ for the envelope, mean $\beta_\mathrm{mm}=1.94\pm 0.12$ for the western core and mean $\beta_\mathrm{mm}=1.84\pm 0.07$ for the eastern core). 
Although \cite{Nguyen-Luong2024} also found a constant $\beta_\mathrm{mm}$ along the transverse profile of the B213 filament in the Taurus molecular cloud, their $\beta_\mathrm{mm}$ was averaged along the filament crest. Therefore, it is hard to compare between our studies. \cite{Kramer2026}, however, implemented a similar strategy as this work to study the Taurus molecular cloud TMC-1. The authors reconstructed the extended emission with \textit{Planck} observations by feathering \citep{Smith2021} and calculated the $\beta_\mathrm{mm}$ map using $T_{Herschel}$. They found spatial variation of $\beta_\mathrm{mm}$ between 1.2 and 1.8 with a typical 1$\sigma$ error of 0.13, which is comparable to our results (Fig.~\ref{fig:beta_mm}) but with a much smaller uncertainty in their case. They argued that the increase of $\beta_\mathrm{mm}$ toward the cloud center is consistent with the modeled increase of grain ice mantle thickness \citep{Ossenkopf1994}, which may also apply in our case, albeit with our large uncertainties. We discuss grain evolution from diffuse type dust to evolved and ice-coated grains in the next subsection.

\subsection{Evidence of dust evolution}\label{sec:dust_evolution}

\begin{figure}[t]
    \centering
    \includegraphics[width=\hsize]{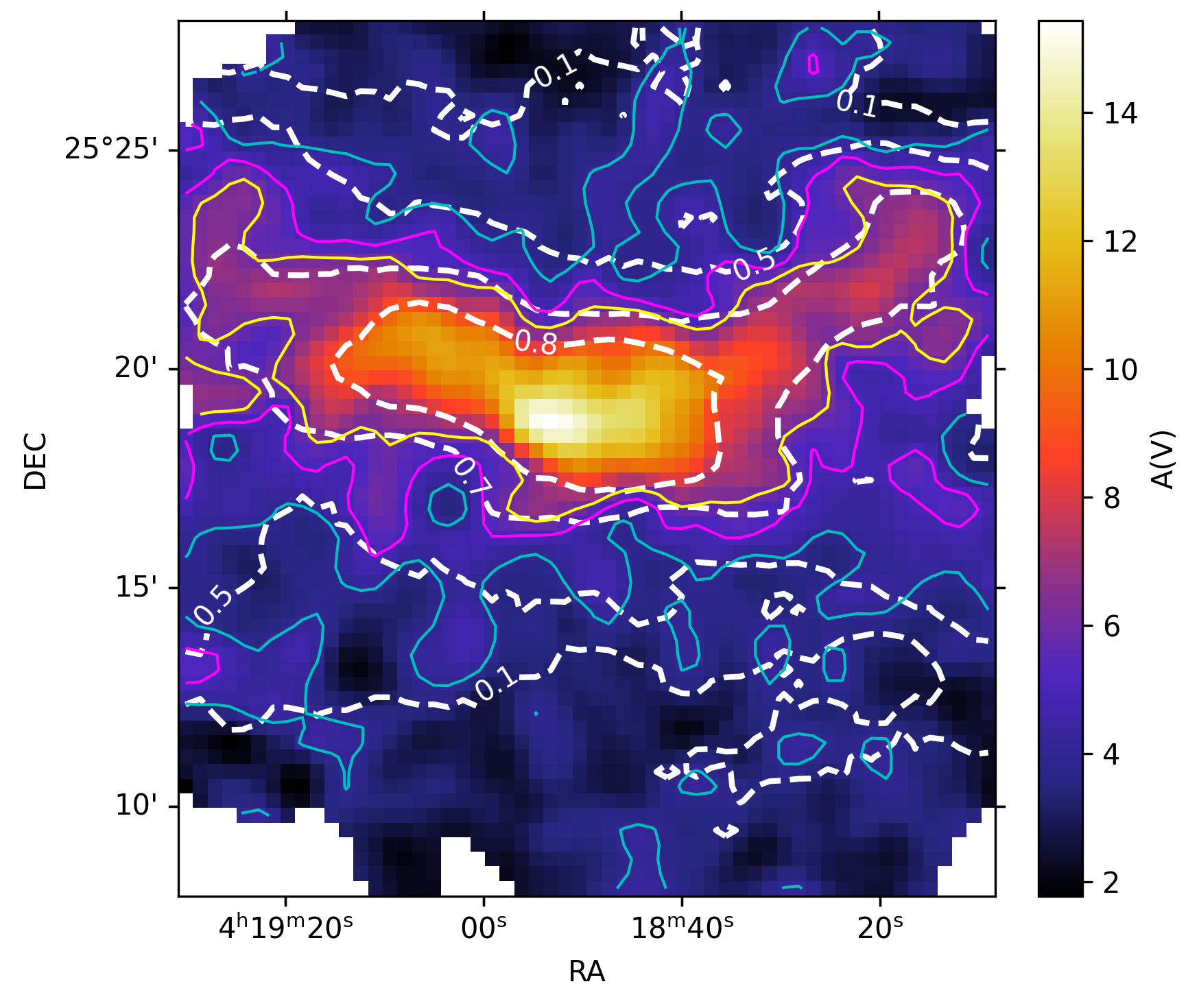}

    \caption{Map of $A_\mathrm{V}$ computed with the NICER method from the WIRCam observations. The white dashed contours show the fraction of dust that, on the LoS, is in evolved grains (at levels of 0.1, 0.5, 0.7, and 0.8). The cyan, magenta and yellow contours show the levels of $A_\mathrm{V}=4,5$, and 6, respectively.} 
    \label{fig:SOC_AV}
\end{figure}

The evolution of the dust properties suggested by the MBB modeling via the spatial variation of $T$ and $\beta$ in the previous subsection is confirmed by the 3D radiative transfer modeling. Despite the degeneracies, our model with two dust components reproduces well the observations (Sect.~\ref{sec:result_SOC}). To confirm this, we also modeled L1506C with only diffuse type dust (the results are presented in Appendix~\ref{app:SOC_diffuse}). The test is motivated by the fact that \citep{Stepnik2003} and \cite{Ysard2013} found that grains more emissive than the diffuse ISM type in the submillimeter are required to reproduce the extinction and emission in L1506C, and THEMIS 2 dust is more emissive than previous diffuse type dust models. The THEMIS 2 dust model we used has $\tau_\mathrm{250}/\tau_\mathrm{J}=7.7\times 10^{-4}$. This value is almost twice the value found by \cite{Planck2013} in diffuse clouds, 1.6 times the value for diffuse type dust model in THEMIS 1 \citep{Jones2013} and more than $2/3$ of the value of the adopted evolved grains ($1\times 10^{-3}$).

The radiative transfer modeling results show that dust emission at 160~$\mu$m is better reproduced with diffuse type dust only, while emission at millimeter wavelengths is better reproduced with two dust populations including evolved grains. This can be explained by the fact that the 160~$\mu$m band traces more the diffuse outer filament, while millimeter wavelengths trace more the denser cores. However, with core residuals $\lesssim20$\% at 2~mm, the dust emission is still reasonably well reproduced using only diffuse type dust, and the extinction is even less overestimated at the same time ($k_J\sim2$ for single dust model, compared to $k_J\sim2.8$ for two dust models, see Table~\ref{table:results_SOC_residuals_diffuse}). As THEMIS 2 dust is more emissive than previous ISM type dust models, SOC is able to reproduce the observed emission and extinction with diffuse type dust only by increasing $n_c$. Residuals in dust emission and extinction are thus not enough to discriminate between the models with one or two dust populations, and we need other constraints, such as $n_\mathrm{H_2}$. If evolved grains are not included in the model, the peak volume density can have values up to $1.5\times10^5$~cm$^{-3}$, higher than derived from independent estimates using molecular lines \citep[N$_2$H$^+$ (1-0),][]{Pagani2010}, which is due to the lower FIR emissivity of diffuse type dust (Fig.~\ref{fig:dust_kappa}). Studies have shown that N$_2$H$^+$ depletion can happen already at $n_\mathrm{H_2}\approx1\times 10^4 \, \mathrm{cm}^{-3}$ in the center of some dense prestellar cores with maximum $n_\mathrm{H_2}>1\times 10^5 \, \mathrm{cm}^{-3}$ \citep[][]{Pagani2007, Lippok2013, Redaelli2019}. If N$_2$H$^+$ is not completely depleted, the central $n_\mathrm{H_2}$ is usually 2-3 times higher than the maximum $n_\mathrm{H_2}$ traced by N$_2$H$^+$ emission. If this is also the case for L1506C, then \cite{Pagani2010} might have underestimated the peak volume density. On the other hand, in their model, \cite{Pagani2010} used the He+N$_2$H$^+$ collisional coefficients \citep{Daniel2005} that are about a factor of two smaller than the H$_2$+N$_2$H$^+$ collisional coefficients documented on the Leiden Atomic and Molecular Database (LAMDA) website\footnote{\url{https://home.strw.leidenuniv.nl/~moldata/}} \citep{Schoier2005}. As a result, their modeled $n_\mathrm{H_2}$ is overestimated, which could overall compensate for the effect of N$_2$H$^+$ depletion.

Another clear signature of grain growth that cannot be reproduced with diffuse type dust only is the phenomenon of coreshine \citep{Andersen2014,Ysard2016}. While the observed level of coreshine can be reproduced using two dust populations (Fig.~\ref{fig:SOC_scattering}), modeling with only diffuse type dust results in $I_\mathrm{sca}$ up to 0.1~MJy/sr that is 2$\sigma$ above the observed value (Fig.~\ref{fig:SOC_scattering_diffuse}). Such a high scattering is unexpected for previous diffuse type dust models. Possible explanation is that THEMIS 2 dust has a high albedo of 0.32 at 3.6~$\mu$m, almost three times larger than the albedo of our evolved grain with $a_0=20$~nm (0.12). \cite{Steinacker2014} also performed radiative transfer modeling of the coreshine observed in this source and conclude that it can be reproduced with a MRN-type dust size distribution with maximal size of 0.65~$\mu$m. It is consistent with our log-normal size distribution of evolved grains with the maximal size of 0.7~$\mu$m (Fig.~\ref{fig:dust_size_dsitribution}). However, we emphasize that grain composition also plays an important role in determining the scattering efficiency. For example, in THEMIS1, due to the second a-C:H mantle, the CMM grains have a higher albedo than the CM grains (Sect.~\ref{sec:dust_model}). Consequently, the same level of scattering can be achieved with dust having smaller sizes \citep{Jones2016, Ysard2016, Ysard2019}. There is not yet an equivalent to CMM grains within the THEMIS 2 framework, but the impact of such grains will be investigated in future work.

Therefore, we conclude that NIR dust extinction and scattering, as well as FIR and millimeter dust emission observations of L1506C can be reproduced with H$_2$ density consistent with measurements from independent molecular tracers only if a population of evolved grains, larger than diffuse ISM with size up to $0.7\;\mu$m, are included. The transition between diffuse type dust and evolved grains occurs within the range: $750 \leq n_0 \leq 2250 \; \mathrm{H_2/cm}^{3}$. For $n_0 =1500\; \mathrm{H_2/cm}^{3}$, evolved grains dominate the dust in the central cores by over 80\% (85\% or 75\% for $n_0=750\; \mathrm{cm}^{-3}$ or 2250~cm$^{-3}$) and in the outer filament by over 50\% (70\% or 40\% for $n_0=750\; \mathrm{cm}^{-3}$ or 2250~cm$^{-3}$) where $A_\mathrm{V} \approx 4$ on the observed extinction map (Fig.~\ref{fig:SOC_AV}).

In the previous radiative transfer modeling of L1506C, \cite{Stepnik2003} found $n_0=1500 \pm 1000$~H$_2$/cm$^{3}$ corresponding to $A_\mathrm{V}=2.1\pm0.5$, and \cite{Ysard2013} found $n_0=750\pm125$~H$_2$/cm$^{3}$, $r_0=0.02$~pc and $n_c=5\times10^4$~H$_2$/cm$^{3}$. Our results agree with theirs regarding $n_0$, and within a factor around two regarding $r_0$ and $n_c$. The differences could be due to several reasons. One is beam dilution: \cite{Stepnik2003} used dust emission observed by the PRONAOS balloon-born telescope with a resolution of 3.5$\arcmin$, almost six times the resolution in our study (36$\arcsec$). \cite{Ysard2013} used extinction from 2MASS data with a spatial resolution of 3.3$\arcmin$, more than three times the resolution in our study (1$\arcmin$). The modeling results also strongly depend on the adopted dust properties, which are not the same in these studies. Our study goes beyond the previous ones, as we modeled a continuous transition between diffuse and evolved grains rather an abrupt transition, which is more realistic. We also modeled the complete cloud, while \cite{Stepnik2003} and \cite{Ysard2013} both modeled individual cross sections.

\subsection{Extinction as a strong constraint for dust properties}\label{discussion: A(J)}

\begin{figure}[tbp]
    \centering
    \includegraphics[width=\hsize]{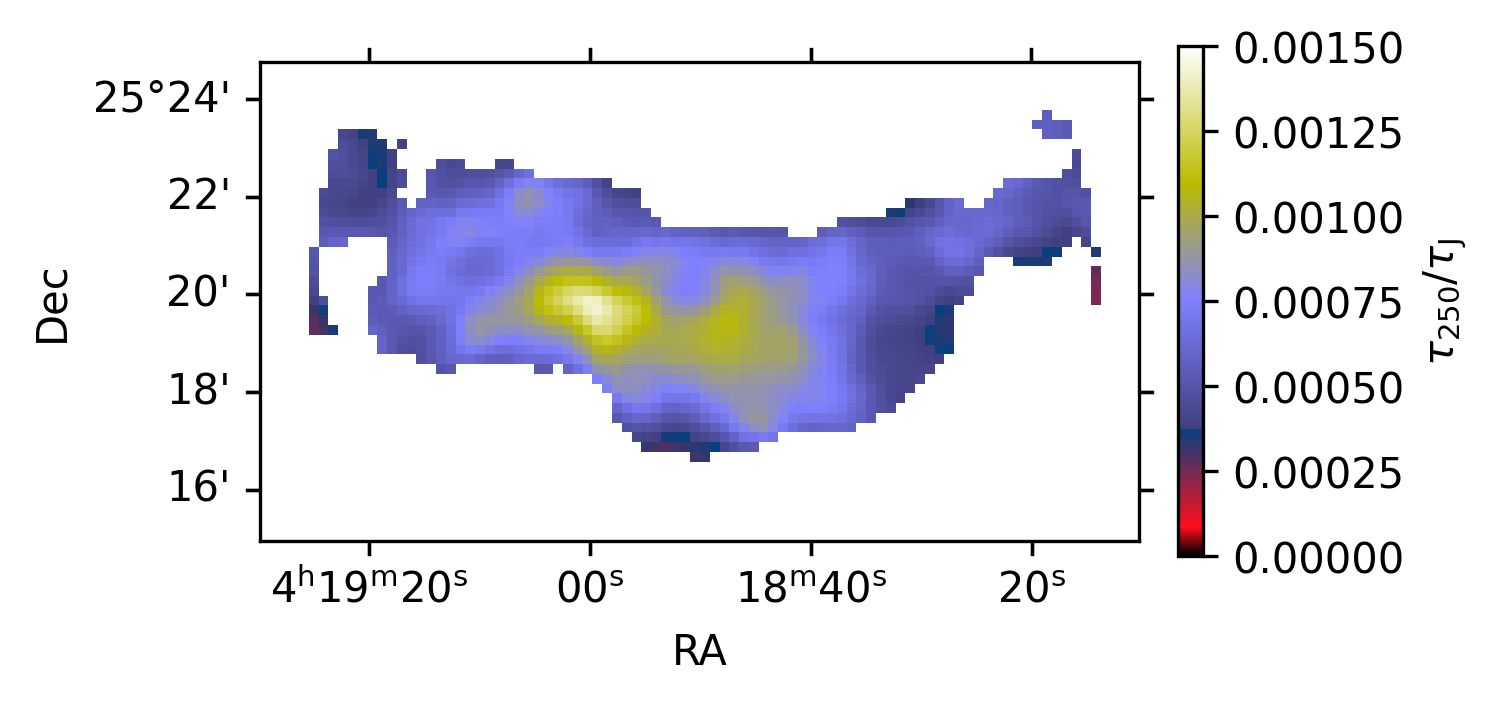}

    \caption{Map of the ratio of $\tau_\mathrm{250}/\tau_\mathrm{J}$ calculated with WIRCam and \textit{Herschel} observations within the MBB fitted area.} 
    \label{fig:tau_map}
\end{figure}

For all parameter sets in our parameter space (Table~\ref{table:SOC_grid}), the modeled $A_\mathrm{J}$ is always overestimated by a factor $k_J \sim2.4-2.8$, even when we exclude the central and densest region of the filament, where the extinction may not be well constrained with WIRCam due to lower star density. This cannot be explained by the geometry of the cloud, because modeling with a cylindrical cloud changes the modeled intensity and extinction by less than 10\%. Instead, the most probable reason for this discrepancy comes from the dust properties. As shown in Fig.~\ref{fig:dust_kappa} and Table~\ref{table:results_SOC_residuals}, decreasing $a_0$ has a noticeable effect on $\kappa$ in the \textit{J}-band and it efficiently lowers $A_\mathrm{J}$. Similar results have been found by \cite{Juvela2024} in his modeling of the Taurus filament LDN~1506 with SOC, where the modeled extinction is also higher by a factor 2$-3$. They noticed that this discrepancy could be reduced, without changing the cloud geometry or radiation field, by increasing $\tau_\mathrm{FIR}/\tau_\mathrm{J}$ by around a factor of three, i.e., by modifying the dust properties (\citealp[see also][for their study of LDN 1642]{Juvela2020} \citealp[and][in Orion]{Juvela2025}). 
We also calculated $\tau_\mathrm{250}/\tau_\mathrm{J}$ of L1506C with the \textit{Herschel} and WIRCam observations. $\tau_{250}$ is calculated as $I_{250}/B_{250}(T)$, where $B_{250}(T)$ is the black body emission at 250~$\mu$m, dependent on the color temperature map derived from MBB modeling. $\tau_J$ is calculated from the WIRCam extinction map, where we used $n_0$ and $a_0$ of the chosen SOC parameter set (in bold in Table~\ref{table:results_SOC_residuals}). The $\tau_\mathrm{250}/\tau_\mathrm{J}$ map is shown in Fig.~\ref{fig:tau_map}. It has a maximum value of $1.4\times 10^{-3}$ within the eastern core area, where evolved grains dominate.
It is higher than $\tau_\mathrm{250}/\tau_\mathrm{J}$ of our adopted evolved grain model ($1\times 10^{-3}$) and of the diffuse type grains ($7.7\times 10^{-4}$).
Overall, these results indicate that dust models having a higher $\tau_\mathrm{FIR}/\tau_\mathrm{J}$ could better reproduce the observed extinction. 
Moreover, given that the $\kappa$ value of silicates varies depending on their composition and structure \citep{Demyk2017}, the $\kappa$ value of THEMIS 2 dust also varies depending on the adopted silicate grain composition \citep{Ysard2024}. In this work, we chose the combination of carbon and silicate grain that is considered the THEMIS 2 reference model \citep{Ysard2024}. A different combination with a higher $\tau_\mathrm{FIR}/\tau_\mathrm{J}$ ratio would influence the modeled results and improve the comparison in extinction.

\section{Conclusion and perspectives}\label{sec:conclusion}

In this work, we have studied the prestellar core L1506C in the Taurus molecular cloud. We made use of the \textit{Herschel} FIR and NIKA2 millimeter observations analyzed with a 3D radiative transfer modeling with SOC that included both diffuse ISM type dust and evolved grains from the recent THEMIS 2 dust model. We combined the \textit{Herschel} and NIKA2 data with a self-developed approach using the software \texttt{scanam\_nika}. For the first time, we performed MBB modeling and 3D modeling of the source over the whole spectral range, from 160~$\mu$m to 2~mm. We also used an NIR extinction map observed with WIRCam and scattering observed with \textit{Spitzer} to constrain the models. The main results are the following:

\begin{itemize}

  \item L1506C is fragmented into two cores, and the MBB modeling revealed their similar properties: a color temperature of $T\approx11$~K, $\beta\approx1.9$, $N_\mathrm{H_2}\approx 2.5 \times 10^{21} \; \mathrm{cm}^{-2}$, and a mass of $0.3\;M_{\odot}$, which is smaller than their Jeans mass.
  \item Both $T$ and $\beta$ within the MBB modeling area show anticorrelated spatial variations beyond uncertainty levels: $T$ decreases from 14.5~K in the envelope surrounding the cores to $\sim11$~K in the cores, and $\beta$ increases from 1.46 in the envelope to $\sim1.9$ in the cores. The anticorrelation has a slope of $-0.06$, which was revealed by the hierarchical Bayesian method. 
  \item The dust emissivity index in NIKA2 wavelength bands ($\beta_\mathrm{mm}$), calculated with the reconstructed NIKA2 maps, has a mean value of $1.88\pm0.27$ within the MBB modeling area with 2~mm weight $>8000$. The value is equal to $2.15\pm0.39$ if one uses the original NIKA2 maps and agrees with other prestellar core values found in the literature.
  \item With 3D radiative transfer modeling, we were able to reproduce the observed millimeter emission with a residual of $\lesssim10$\% for the cores and coreshine within $1\sigma$. Extinction is still overestimated by a factor $\sim2.8$. The cloud geometry is degenerate and cannot be well constrained with only dust emission or NIR extinction data.
  \item Radiative transfer modeling revealed in 3D a central dust temperature as low as 8~K, which is about 3~K lower than derived from MBB modeling. The model $N_\mathrm{H_2}$ has a maximum value of $1.3 \times 10^{22}$~cm$^{-2}$, about 1.7 times higher than the one derived from MBB modeling, indicating the importance of radiative transfer modeling for correctly constraining $N_\mathrm{H_2}$.
  \item The modeling of the radiative transfer shows that porous ice-coated grains with a size of up to 0.7~$\mu$m, which is larger than those found in the diffuse interstellar medium, are necessary to adjust the FIR-millimeter emission and to reproduce the coreshine effect while having a cloud volume density compatible with measurements from independent observations of molecular lines.
  \item The grain evolution from diffuse type to evolved grains occurs at densities ($n_\mathrm{H}$) between 1500 and 4500~$\mathrm{cm}^{-3}$. For a threshold density of $n_0=$3000~$\mathrm{H/cm}^{3}$, the amount of evolved grains reaches 50\% at $A_\mathrm{V} \approx 4$.
  \item Observations of extinction impose significant constraints on the models. Using dust with different properties, particularly a lower $a_0$ and a higher optical depth ratio between the FIR and \textit{J} bands, could reduce the discrepancy between modeled and observed extinction.   
\end{itemize}

We will extend this work to other prestellar and protostellar cores with different evolutionary stages located at various galactic positions to have a comprehensive view of grain evolution at early stages of star formation. 
We will also explore different grain properties by varying the grain composition within the framework of THEMIS 2 dust models.

\begin{acknowledgements}
       We thank the referee for their careful reading of the manuscript and constructive comments. This work was supported by the Thematic Action ``Physique et Chimie du Milieu Interstellaire'' (PCMI) of INSU Programme National ``Astro'', with contributions from CNRS Physique \& CNRS Chimie, CEA, and CNES; and through the grant EUR TESS N°ANR-18-EURE-0018 in the framework of the Programme des Investissements d'Avenir. We would like to thank the IRAM staff for their support during the observation campaigns. The NIKA2 dilution cryostat has been designed and built at the Institut N\'eel. In particular, we acknowledge the crucial contribution of the Cryogenics Group, and in particular Gregory Garde, Henri Rodenas, Jean-Paul Leggeri, Philippe Camus. This work has been partially funded by the Foundation Nanoscience Grenoble and the LabEx FOCUS ANR-11-LABX-0013. This work is supported by the French National Research Agency under the contracts "MKIDS", "NIKA" and ANR-15-CE31-0017 and in the framework of the "Investissements d’avenir” program (ANR-15-IDEX-02). This work has been supported by the GIS KIDs. This work has benefited from the support of the European Research Council Advanced Grant ORISTARS under the European Union’s Seventh Framework Programme (Grant agreement No. 291294).
       The NIKA2 data processing with {\it Scanam\_nika} was done on the Infinity Cluster hosted by Institut d'Astrophysique de Paris. We thank St\'ephane Rouberol for smoothly running this cluster and Christophe Pichon for managing its use and upgrades.
       We acknowledge the use of data and softwares provided by the Centre d'Analyse de Données Etendues (CADE), a service of IRAP-UPS/CNRS \citep[\url{http://cade.irap.omp.eu},][]{Paradis2012}. MJ acknowledges the support of the Research Council of Finland Grant No. 348342.
\end{acknowledgements}

\bibliographystyle{aa}
\bibliography{reference}

\begin{appendix}

\section{NIKA2 data processing}\label{appendix:nika2}

\subsection{Noise subtraction}

The subtraction of atmospheric and instrumental noise from the NIKA2 raw calibrated data is performed with the \textit{Scanam\_nika} map-making tool \citep{Roussel2013}, following these principles:

\begin{itemize}
\item The algorithm exploits the redundancy built in the observations to separate the noise from the astrophysical signal, each position on the sky being sampled by many detectors, at many different times spread over the whole observing period, and with two distinct scan orientations, each chosen to make as large an angle as possible to the main elongated structure in the cloud as well as to the other scan orientation.

\item The noise is decomposed into several categories, tackled in order of both decreasing correlation between detectors and decreasing amplitude, on varying timescales. The linear part of the noise can be easily subtracted as baselines fitted to the time series, while the nonlinear part is subtracted by exploiting the redundancy within each cell of the spatial grid adapted to that problem, determined by the sampling rate and the scan speed.

\item We subtract first the average drift (mostly corresponding to the atmospheric emission), defined as a single function of time for the whole 2\,mm array and a single function of time for the set of both 1.15\,mm arrays, on the shortest accessible timescale; then the median noise per electronic box (four functions of time at 2\,mm and sixteen functions of time at 1.15\,mm), on successively larger timescales, from the 2\,mm FWHM crossing time up to about half the map crossing time; then the residuals of the correlated electronic noise (subtracted by filtering as explained in Ejlali et al. 2026, in prep.); then the uncorrelated noise (a function of time for each individual detector), on the same timescales but in the reverse order.

\item To avoid subtracting astrophysical signal, a source mask is rebuilt at each important step of the processing. We use a prior, defined as the area enclosed within the 11.1~MJy/sr isophote in the 500~$\mu$m map, to define the maximal extent of the mask and prevent it from covering areas with strong positive noise. Within that area, the background is anchored to pixels with the lowest brightness, and linearly interpolated in the other pixels. The difference between the actual map and the background map is then divided by the error map to obtain a S/N map. The source mask is based on it, using evolutive thresholds as the noise progressively decreases.
\end{itemize}

\subsection{Reconstruction of extended emission}

\begin{figure}[htbp]
    \includegraphics[width=\hsize]{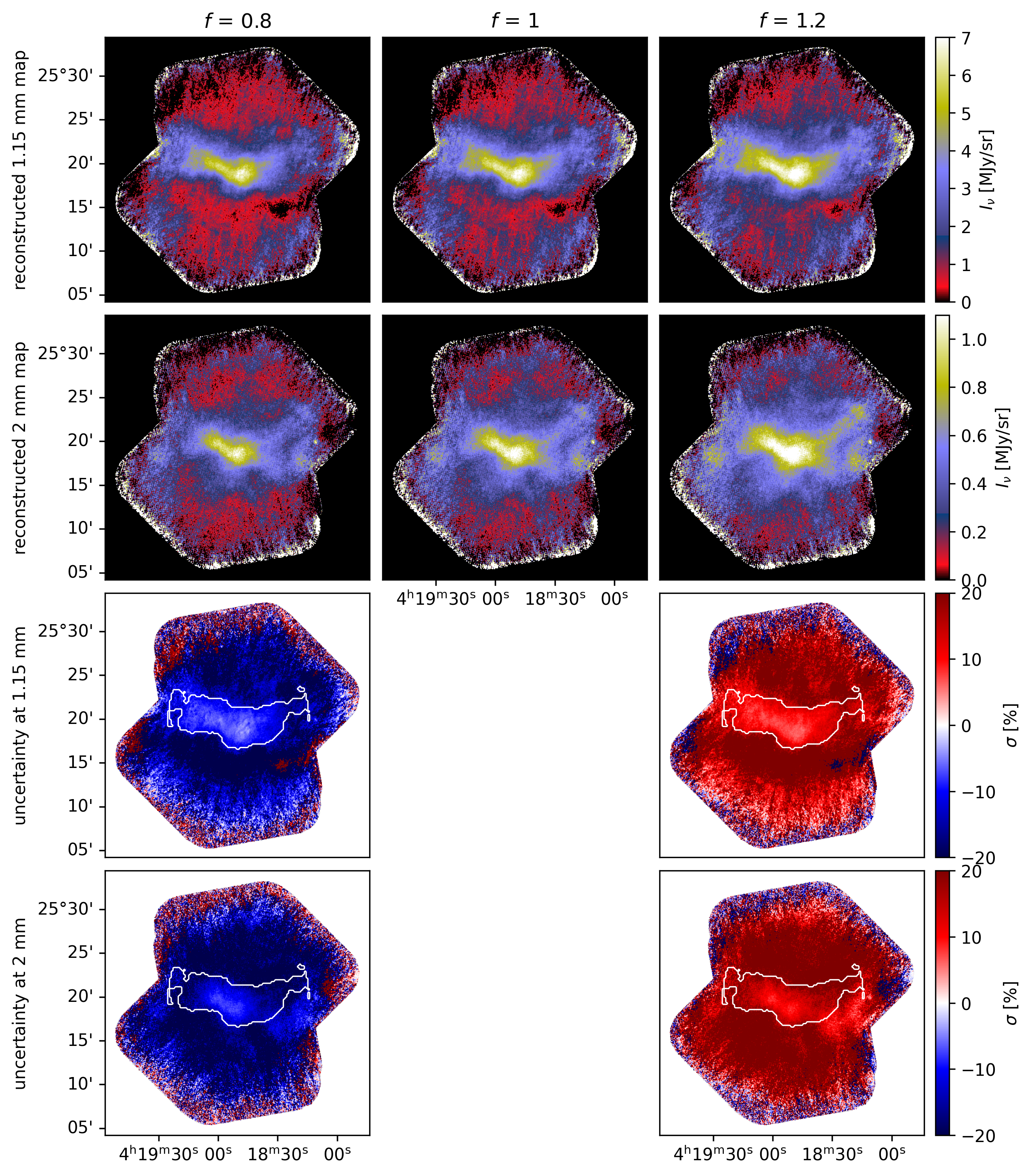}

    \caption{Reconstructed NIKA2 maps of different factor $f$. The columns from left to right show intensity maps with $f=0.8$, $f=1$ and $f=1.2$ respectively ($I_{f=0.8}$, $I_{f=1}$, $I_{f=1.2}$). The top row shows maps at 1.15~mm with a resolution of 12$\arcsec$ and the second row shows maps at 2~mm with a resolution of 18$\arcsec$. The third and bottom row show uncertainty of the reconstructed maps of $f=1$ at 1.15 or 2~mm, calculated from the maps of $f=0.8$ or $f=1.2$ (e.g., (  $I_{f=0.8} -I_{f=1}$ ) / $I_{f=1}$ for the lower uncertainty). The white contours on the bottom two rows show the MBB modeling area.} 
    \label{fig:comparison_factor}
\end{figure}

To preserve extended emission while processing the NIKA2 data, we use prior information from the predictions based on \textit{Herschel} and \textit{Planck} maps (Sect.~\ref{sec:data_reduction}).

The predicted maps are prone to some errors, and we therefore did not impose them as the truth for extended emission. The principle is instead to subtract the model from the calibrated NIKA2 data, leaving on top of the atmospheric and instrumental noise mostly small-scale emission (suffering much less filtering than large-scale emission), to process these data with the default method, and at the end to reinject the model initially subtracted.

We assume that a 20\% uncertainty is conservative at 1.15\,mm and realistic at 2\,mm, since \cite{Paradis2024} report a 7\% uncertainty at 1.38\,mm, in a range of environments encompassing those of cold cores. Reaching 2\,mm requires an extrapolation beyond the wavelength domain where the neural network model was extensively tested, hence the larger uncertainty.

New time series are generated as follows:
\begin{equation}
    S_\mathrm{new} = S_\mathrm{NIKA2} - f \times S_\mathrm{predic}
\end{equation}
where $S_\mathrm{NIKA2}$ designates the raw time series (corrected only for the various multiplicative effects) and $S_\mathrm{predic}$ the mock time series simulated from the predicted map at 1.15 or 2\,mm.
The default value is $f=1$, but values of $f=0.8$ and $f=1.2$ were also used to estimate the final uncertainties stemming from a 20\% error bar for the predictions.

$S_\mathrm{new}$ were then processed with \texttt{Scanam\_nika} in the same way as the original NIKA2 data, as outlined in Sect.~\ref{sec:scanam} to produce a map
$S_{\rm new, processed}$. The only difference is that both positive and negative sources have to be allowed to build the source mask at each step, because the models can produce either under- or overpredictions depending on the location, and because they have a coarser angular resolution: even for a perfect model, subtracting a smoothed version of a compact source leaves a positive excess at the center but a negative annulus around.

At the end of the data processing, the subtracted intensity was reinjected before the final projection on the sky:
\begin{equation}
    S_\mathrm{recon} = S_\mathrm{new,\;processed} + f \times S_\mathrm{predic}
\end{equation}

The reconstructed maps at their nominal resolution of different factor $f$ are shown in Fig.~\ref{fig:comparison_factor}. The larger the factor is, the more extended emission is reconstructed and the intensity increases. We calculated the uncertainty of the maps of $f=1$ from maps of other factors (e.g., (  $I_{f=0.8} -I_{f=1}$ ) / $I_{f=1}$ for the lower uncertainty). Within the MBB modeling area where S/N$\geq 4$ at all wavelength bands (including reconstructed NIKA2 maps of $f=1$), the intensity uncertainty is $\leq10$\% for the majority of the pixels, showing the robustness of the reconstruction method.

\section{Background determination}\label{app:bkg}

\begin{figure}[htbp]
    \centering
    \includegraphics[width=\hsize]{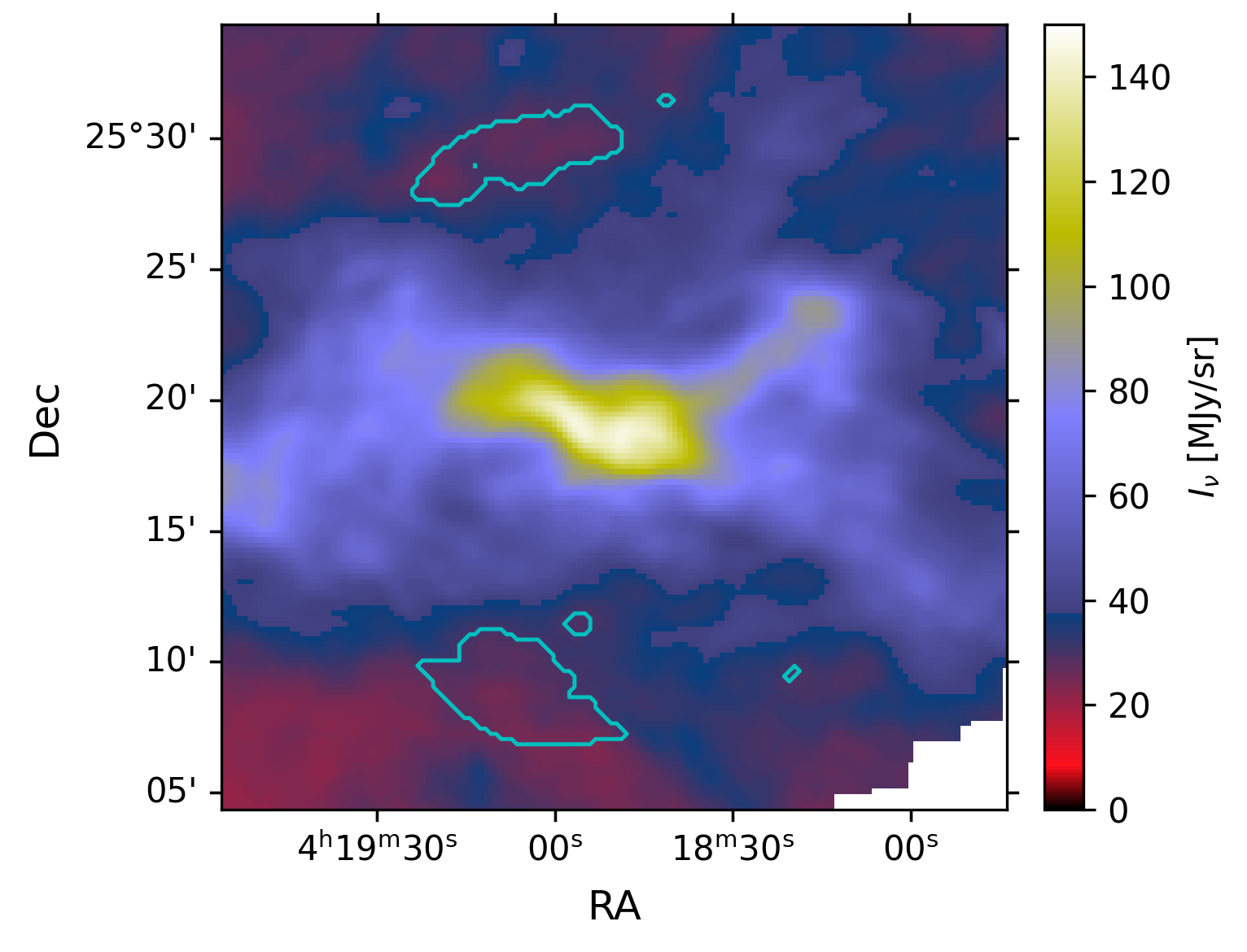}
    \includegraphics[width=\hsize]{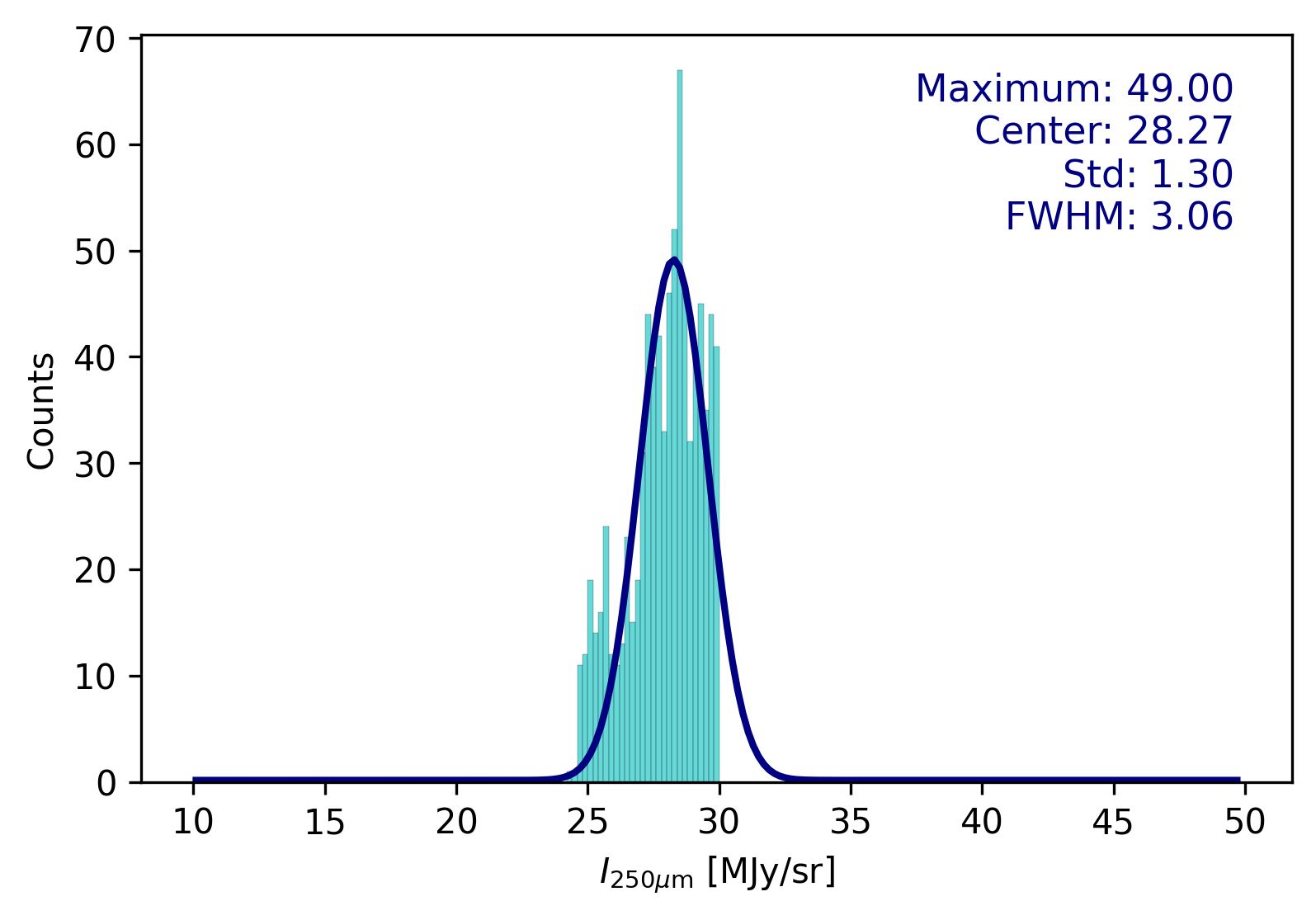}

    \caption{Area to determine the background and the background histogram. \textit{Top panel}: \textit{Herschel} 250~$\mu$m map resampled to 12$\arcsec$ pixel size and smoothed to 36$\arcsec$ beam resolution. The cyan contours show regions within which the background was determined. \textit{Bottom panel}: Histogram of $250$~$\mu$m pixel intensity within the background region and the fitted Gaussian function. Maximum, center intensity, standard deviation and FWHM of the function are shown on the top right.} 
    \label{fig:bkg_mask}
\end{figure}

Figure~\ref{fig:bkg_mask} shows the region within which the background is determined ($I_{250} <30$~MJy/sr and weight of the 1~mm intensity map $>30000$), and the corresponding histogram and fitted Gaussian function.

\section{Flux uncertainty of the NIKA2 maps}

\begin{figure}[htbp]
    \centering
    \includegraphics[width=\hsize]{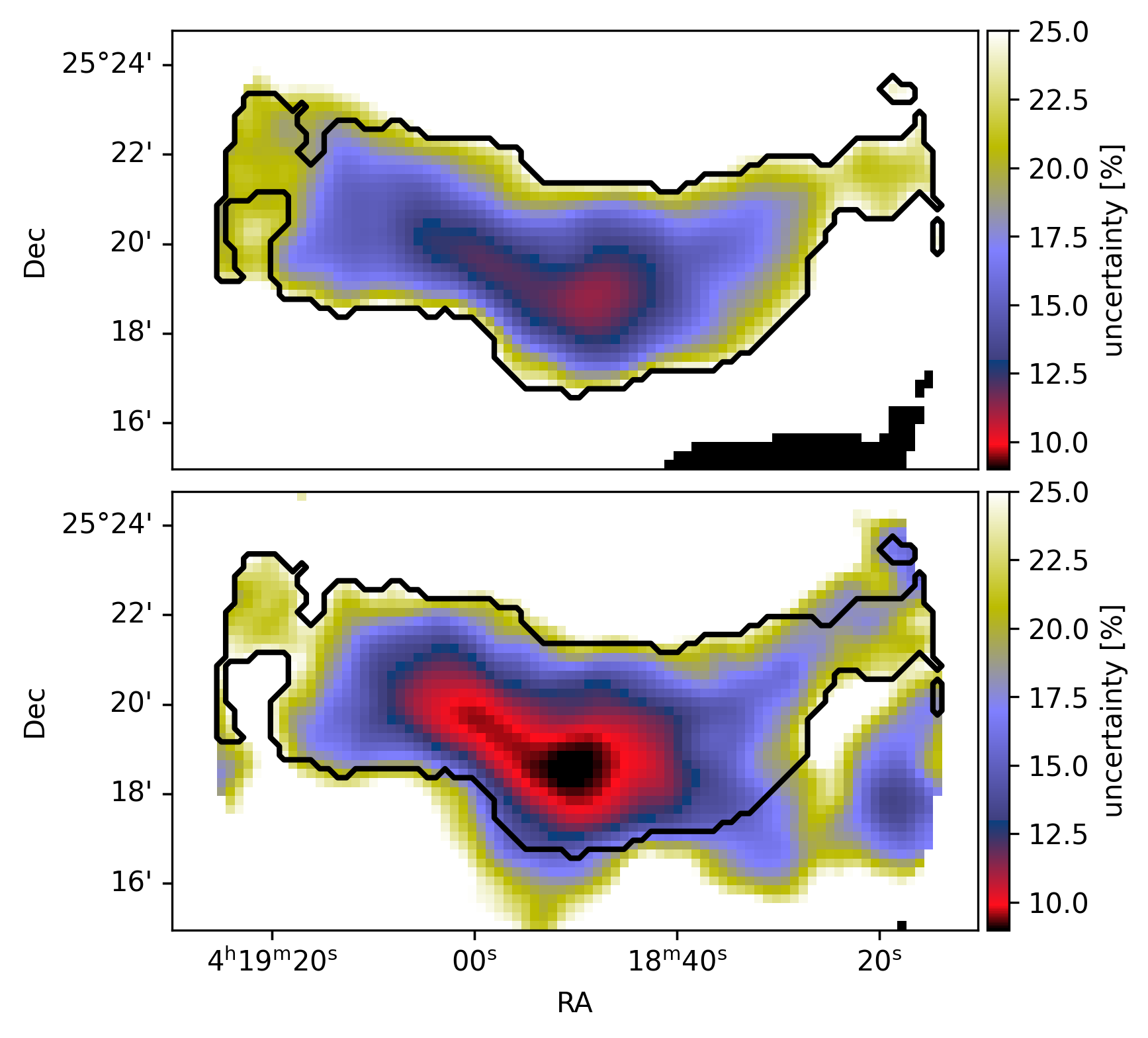}

    \caption{Flux uncertainties in percentage of the NIKA2 1.15~mm map (top panel) and 2~mm map (bottom panel). The black contours show the MBB fitted area.} 
    \label{fig:map_uncertainty}
\end{figure}

Flux uncertainties of the NIKA2 maps within the MBB fitted area are shown in Fig.~\ref{fig:map_uncertainty}. The uncertainty calculation method is presented in Sect.~\ref{Sec: method_maps}.

\section{Hierarchical Bayesian fit results}\label{app:hierarchical_bayesian}

\begin{figure*}
\sidecaption
  \includegraphics[width=12cm]{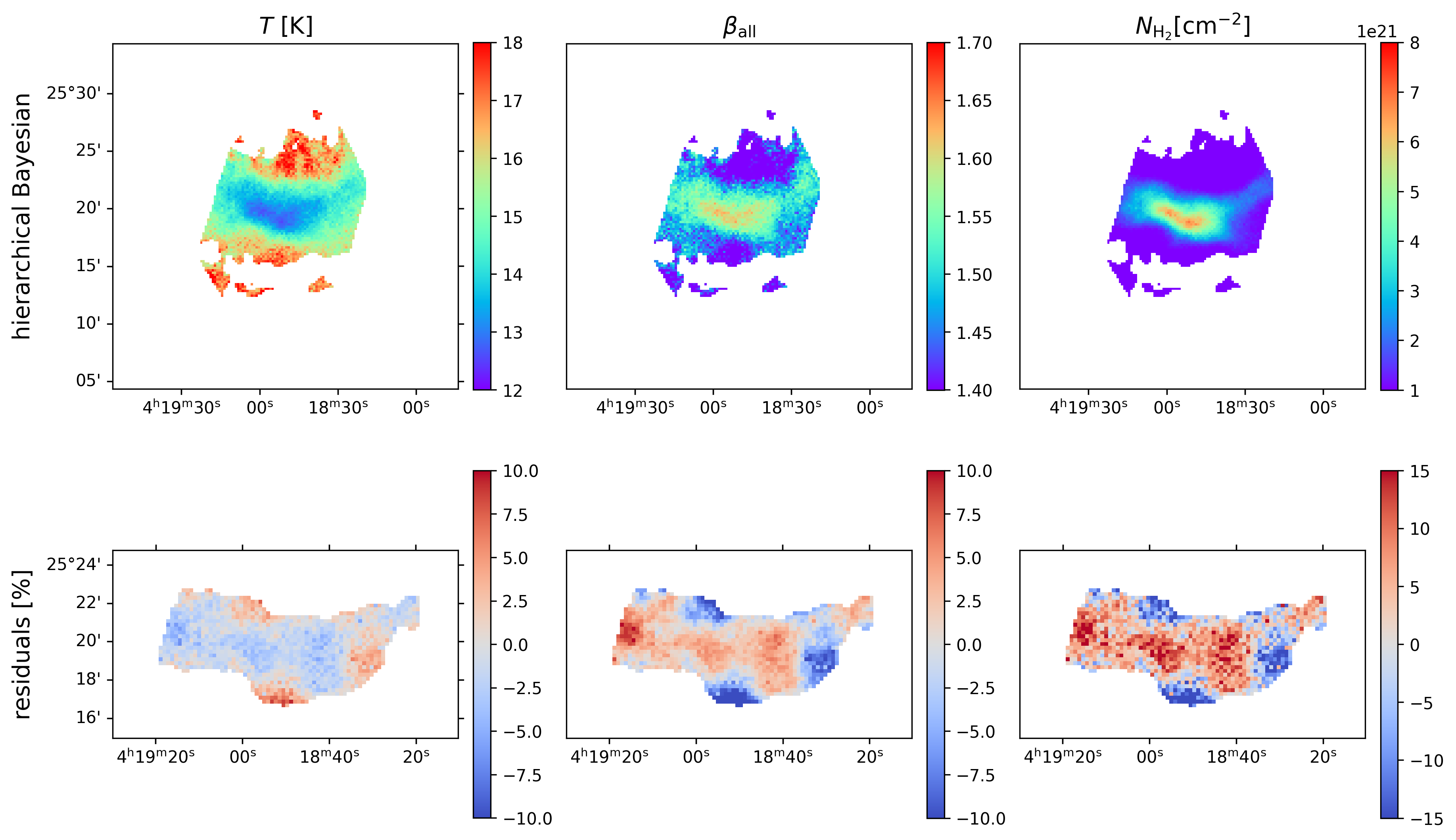}
     \caption{MBB fit maps with hierarchical Bayesian method and comparison with maps with $\chi^2$ minimization method. The top row shows the maps derived with a hierarchical Bayesian method, the bottom row shows the residual in percentage (($\chi^2$ minimization map $-$ hierarchical Bayesian map)/$\chi^2$ minimization map). The three columns from left to right show $T$, $\beta_\mathrm{all}$ and $N_\mathrm{H_2}$ respectively.}
     \label{fig:MBB_bayesian}
\end{figure*}

We present here the results of the MBB modeling performed using a hierarchical Bayesian method (Fig.~\ref{fig:MBB_bayesian}). We assumed that the prior distribution and flux probability both follow a Student t-distribution. Adopting the Gaussian distribution instead of the Student t-distribution varies $T$ and $\beta$ by no more than 5\% in the central part where S/N $\ge 4$ for the majority of the pixels. Compared to the results obtained with the $\chi^2$ minimization, the results obtained with the hierarchical Bayesian method have a smaller spread: higher $T$, smaller $\beta$ and smaller $N_\mathrm{H_2}$ in the center, and lower $T$, higher $\beta$ and higher $N_\mathrm{H_2}$ at the outskirts (Fig.~\ref{fig:MBB_bayesian}). However, the difference is about 10\% or less for $T$ and $\beta$, and about 15\% or less for $N_\mathrm{H_2}$, which confirms that the results of $\chi^2$ minimization are valid. Note that the fitted area for the hierarchical Bayesian method is larger than the one for $\chi^2$ minimization, but in Fig.~\ref{fig:T_beta_mm} the anticorrelation is plotted only for data points with S/N $\ge 4$ at all wavelengths.

To compare with the aperture photometry results obtained using $\chi^2$ minimization presented in Sect.~\ref{sec:result_core_properties}, we also did a MBB fit with the hierarchical Bayesian method for pixels within the cores area after subtracting the envelope intensity (listed in Table~\ref{table:SED_values}) from each pixel. The resulting mean $T$ and $\beta$ within the two cores area is 11.0~K and 1.90 respectively. Within the envelope area, where only the background is subtracted, we obtain mean $T=14.4$~K and mean $\beta=1.47$. These values agree with the results presented in Sect.~\ref{sec:result_core_properties} ($T=10.7\pm0.2$~K and $\beta =1.95\pm0.07$ for the eastern core, $T=11.3\pm0.2$~K and $\beta =1.83\pm0.05$ for the western core, and $T=14.5\pm0.7$~K and $\beta=1.46\pm0.11$ for the envelope).

\FloatBarrier

\section{Cores, envelope and filament SEDs}

\begin{table}[h]
\caption{Intensity and 1$\sigma$ uncertainty in $\mathrm{MJy\,sr}^{-1}$ at different wavelengths of the cores, envelope and filament SED shown in Fig.~\ref{fig:core_SED}.}

\label{table:SED_values}  
\centering       

\setlength{\tabcolsep}{3pt}
\begin{tabular}{c|cccc}  
$\lambda$ [$\mu$m] & eastern core & western core & envelope & filament\\
\hline 
\hline
160 & 19.02$\pm$2.55 & 24.04$\pm$2.75 & 42.01$\pm$3.66 & 31.34$\pm$3.09 \\
250 & 45.73$\pm$3.45 & 47.98$\pm$3.60 & 58.60$\pm$4.30 & 37.65$\pm$2.94 \\
350 & 40.34$\pm$2.94 & 42.07$\pm$3.05 & 41.07$\pm$2.99 & 24.46$\pm$1.89 \\
500 & 21.74$\pm$1.57 & 22.90$\pm$1.65 & 19.67$\pm$1.43 & 11.27$\pm$0.88 \\
1150 & 2.20$\pm$0.52 & 2.61$\pm$0.53 & 2.78$\pm$0.53 &  \\
2000 & 0.35$\pm$0.07 & 0.37$\pm$0.07 & 0.42$\pm$0.07 &  \\
850 &               &                 &             & 3.23$\pm$0.16 \\
1400 &             &                 &             &  0.76$\pm$0.07  \\

\end{tabular}
\tablefoot{Because the filament has low S/N on the NIKA2 maps, its SED is instead built using the predicted \textit{Planck} maps.}
\end{table}

The intensities and their uncertainties of the cores, envelope and filament SEDs plotted in Fig~\ref{fig:core_SED} are listed in Table~\ref{table:SED_values}. The uncertainties were calculated as the quadratic sum of background uncertainties and of the calibration uncertainties, as explained in Sect.~\ref{Sec: method_maps}.

\FloatBarrier

\section{Cloud geometry from SOC modeling}\label{app:SOC_n}

\begin{figure}
    \centering
    \includegraphics[width=7.5cm, trim=1.5cm 2.5cm 1.5cm 2.5cm,
    clip]{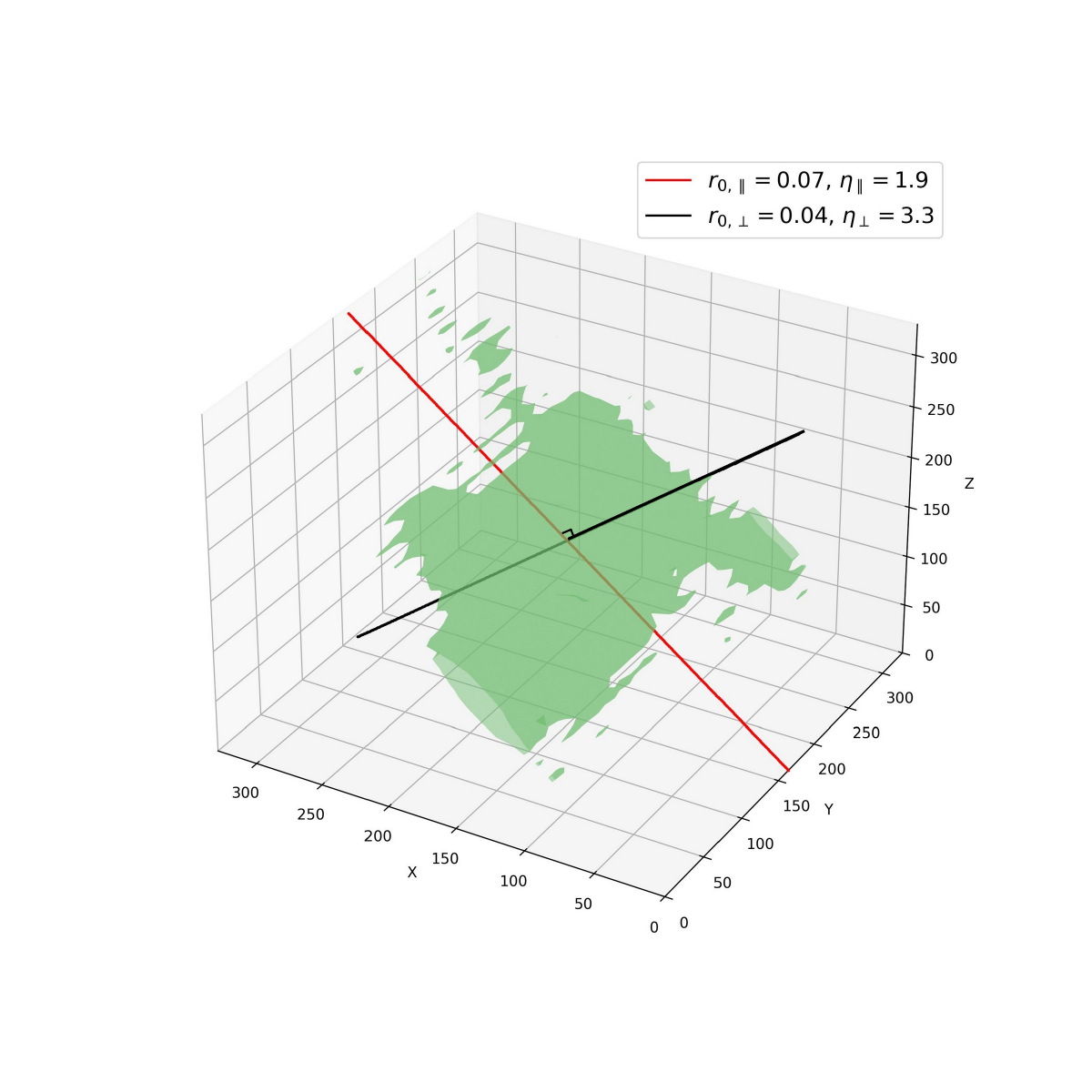}
    \includegraphics[width=6.5cm]{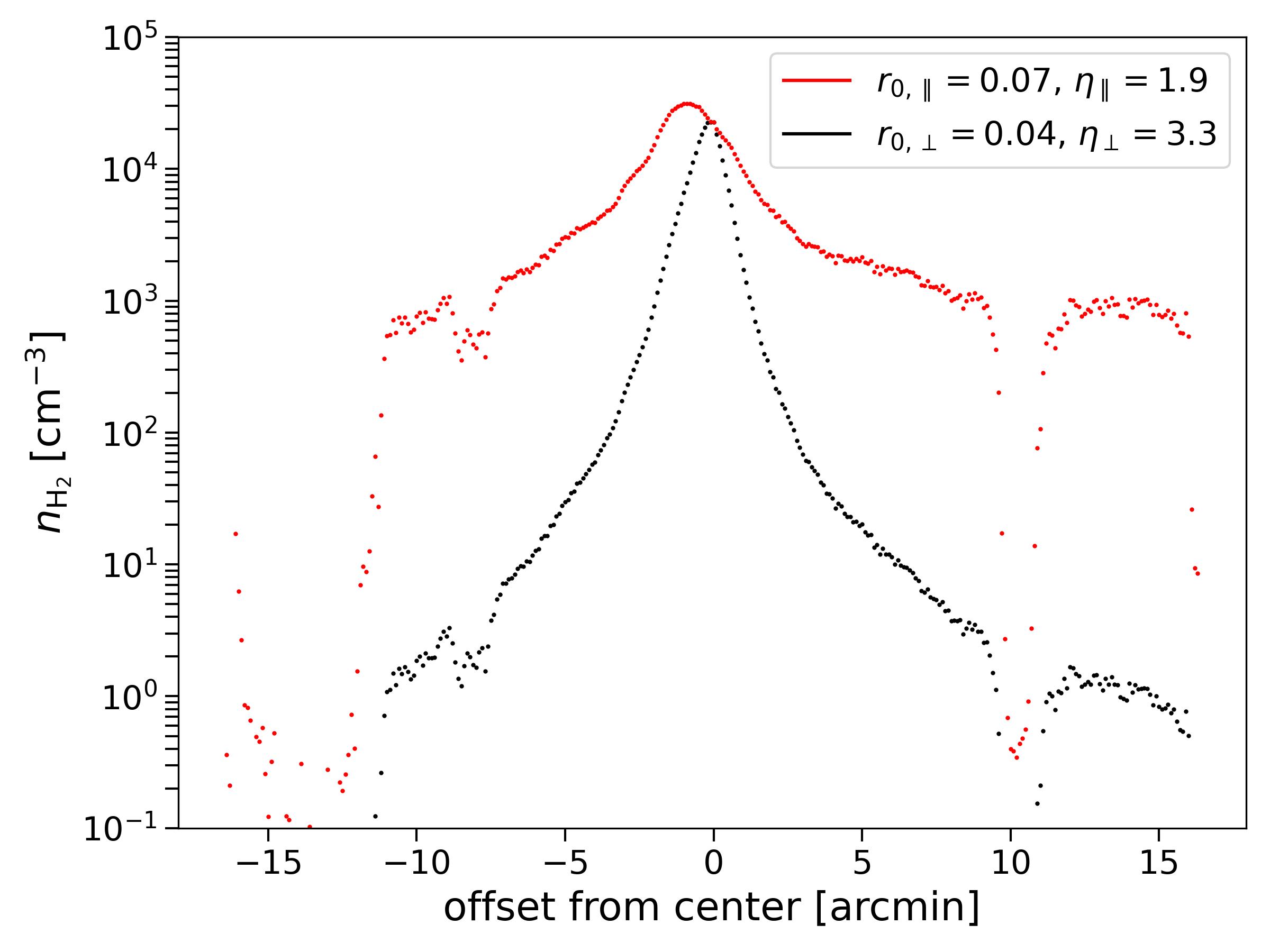}

    \caption{The modeled L1506 density structure. \textit{Top panel:} isosurface of level $1500 \; \mathrm{cm}^{-3}$ of the SOC modeled $n_\mathrm{H_2}$ of the chosen parameter set in boldface in Table~\ref{table:results_SOC_residuals}. The red and black perpendicular solid lines show the directions along which we calculated the density profiles parallel and perpendicular to the filament plane. The LoS direction is the z-axis, right ascension corresponds to the y-axis and declination corresponds to the x-axis. \textit{Bottom panel:} $n_\mathrm{H_2}$ density profiles along the parallel and perpendicular directions shown on the left panel. The x-axis shows offset in arcminutes along each profile relative to the center of the modeled box.} 
    \label{fig:SOC_nH}
\end{figure}

Figure~\ref{fig:SOC_nH} shows the 3D isosurface of level $1500 \; \mathrm{cm}^{-3}$ of the SOC modeled $n_\mathrm{H_2}$ for the chosen parameter set with two dust populations (in bold in Table~\ref{table:results_SOC_residuals}). The cloud looks like an inclined flattened cylinder with an inclination angle of 45$^\circ$. The inclination is applied artificially when the LoS structure is defined. The degree of inclination changes the modeling results by less than 5\% for all parameters except $G_0$, $r_{0, \perp}$, $\eta_\perp$ and the core residual at 160~$\mu$m, for which the changes can be up to 30$-$50\%. However, these parameters are not well constrained and do not change the conclusions of our analysis. We corrected for inclination when calculating the cloud density profiles so that $r_{0, \parallel}$ and $\eta_\parallel$ belong to the profile along the filament plane rather than the plane of the sky (x-y plane), and the perpendicular direction is perpendicular to the filament plane rather than simply the z-axis.

\FloatBarrier

\section{Dust models and properties}

\begin{figure}[htbp]
    \centering
    \includegraphics[width=\hsize,trim={0.5cm 0cm 1.5cm 1cm}, clip]{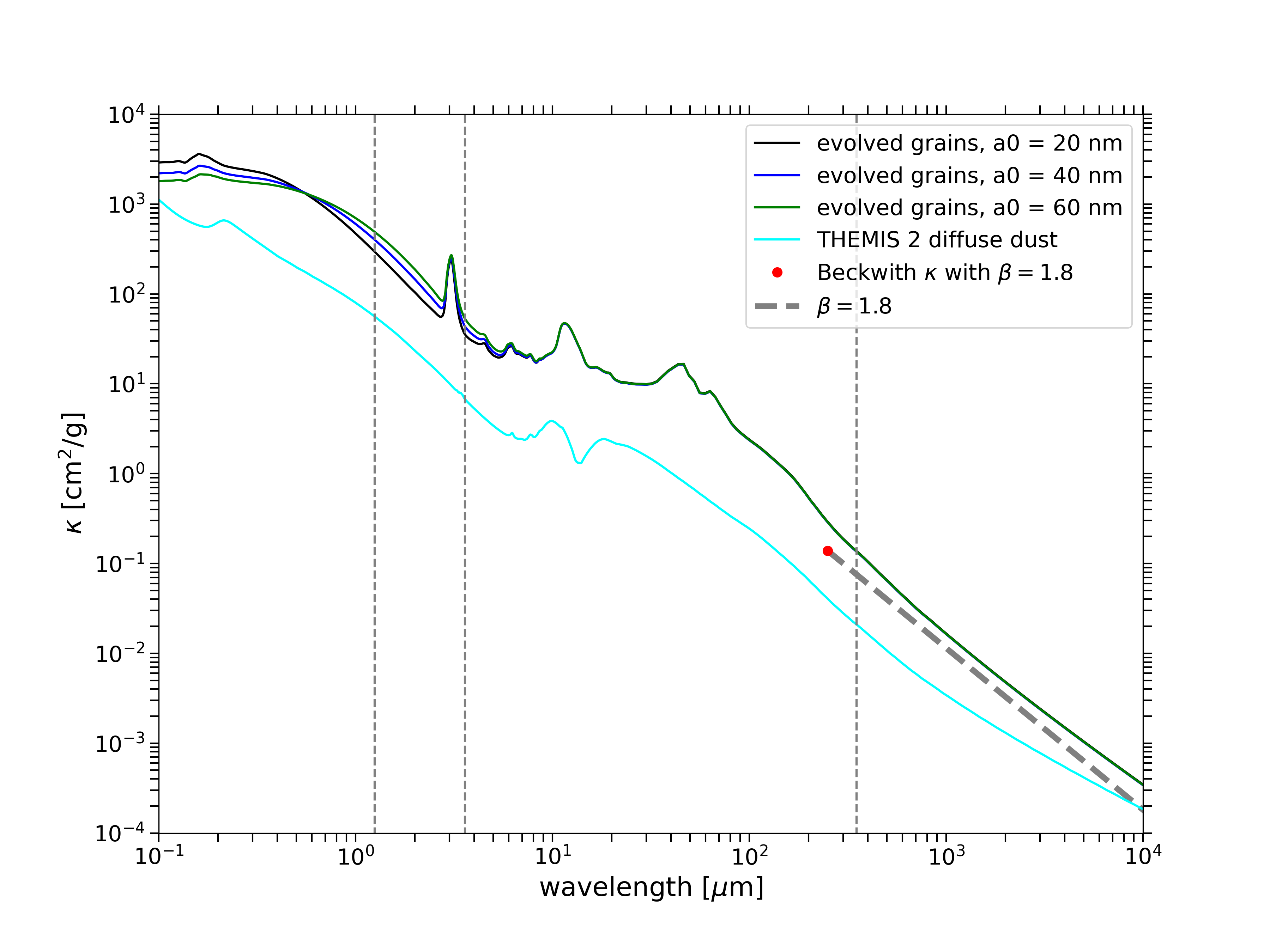}

    \caption{Mass extinction coefficient, $\kappa$, of the dust models used in this study. Cyan line: $\kappa$ of THEMIS 2 diffuse type grains. Black, blue and green lines: $\kappa$ of evolved grains calculated from the THEMIS 2 diffuse type dust for size distributions with  a$_0$ of 20, 40 and 60~nm, respectively. The red dot shows B90 $\kappa$ at 250~$\mu$m and the thick dashed gray line shows extrapolation with $\beta=1.8$. The thin dash-dotted and the dotted lines show the function $\kappa_\nu=\kappa_0(\nu/\nu_0)^\beta$, with $\nu_0$ at 1~mm and $\beta=1.8$ or 2.2 for THEMIS 2 dust or evolved grains respectively. The vertical dashed line shows position of the $J$ (1.25~$\mu$m), 3.6 $\mu$m and 350 $\mu$m bands.} 
    \label{fig:dust_kappa}
\end{figure}

\begin{figure}[htbp]
    \centering
    \includegraphics[width=\hsize, trim=0cm 1.5cm 0cm 0cm, clip]{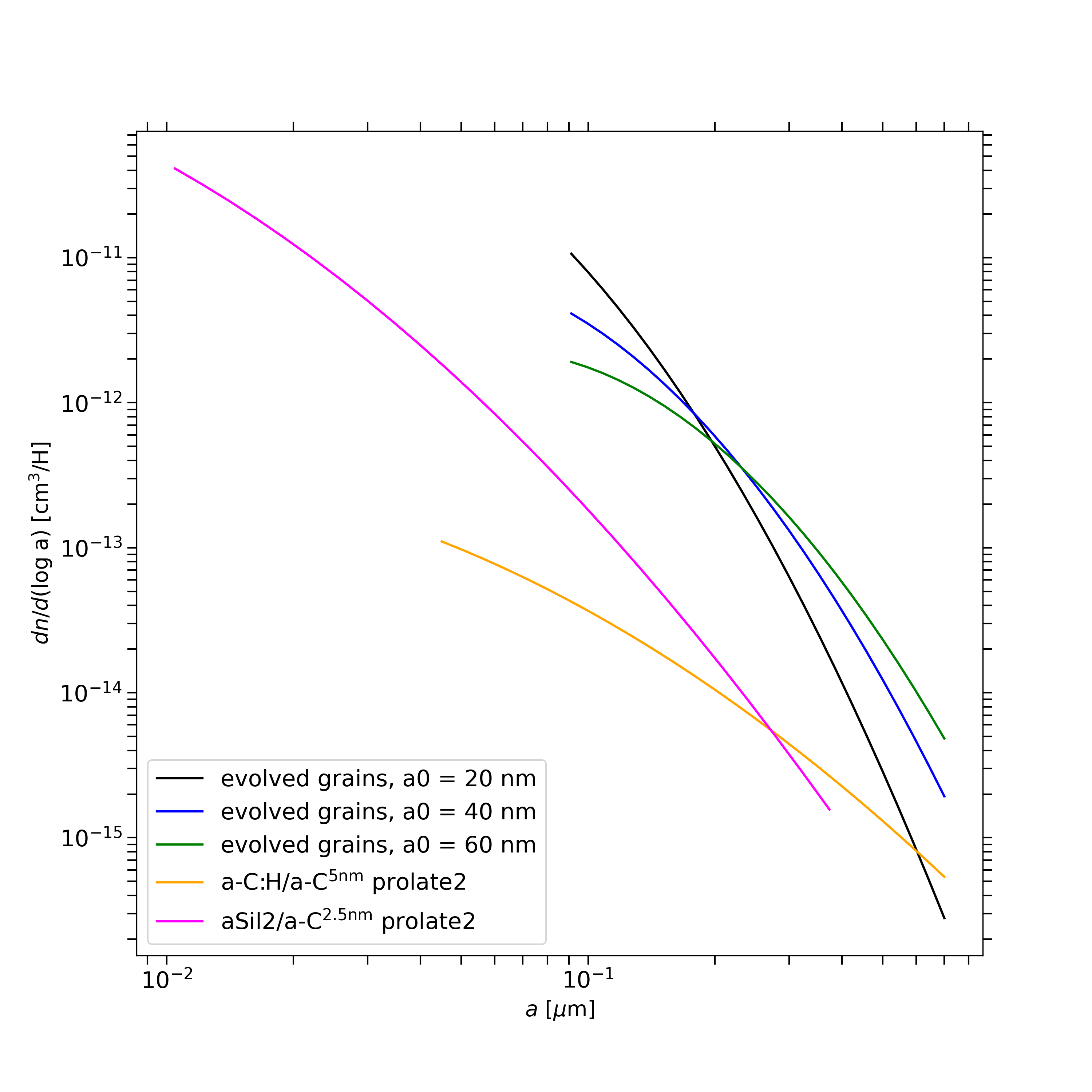}
    \includegraphics[width=\hsize, trim=0cm 0cm 0cm 1.5cm, clip]{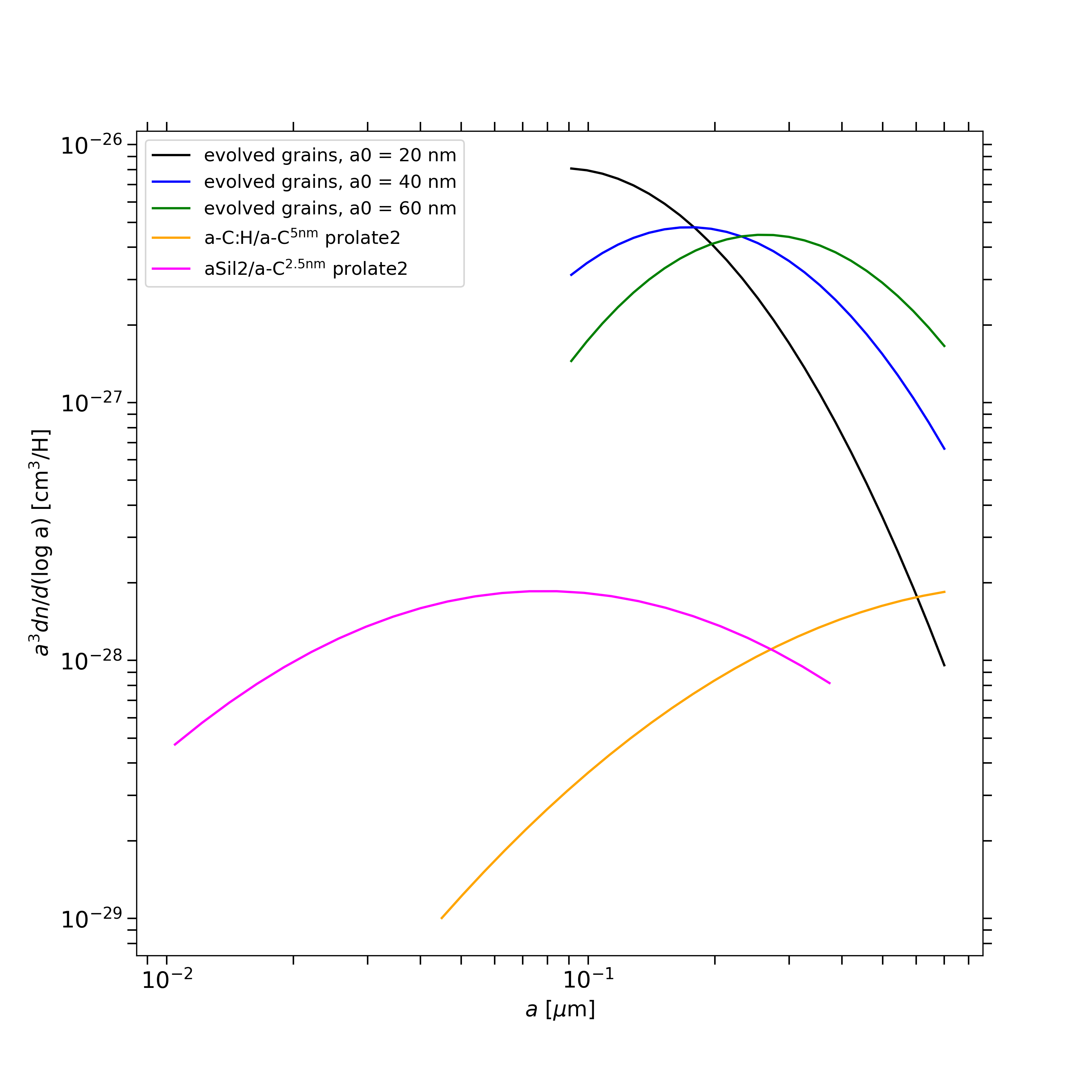}

    \caption{Dust size distribution (top) and mass distribution (bottom) of THEMIS 2 diffuse dust and evolved grains. 
    The magenta and orange curves show the two dust components of THEMIS 2 dust: silicate and carbon grains respectively. The distributions of evolved grains for different values of $a_0$ are also plotted (black, blue and green curves).} 
    \label{fig:dust_size_dsitribution}
\end{figure}

The mass extinction coefficient and size distribution of the dust models used in the SOC modeling are shown in Figs.~\ref{fig:dust_kappa} and ~\ref{fig:dust_size_dsitribution}. We also fitted $\kappa$ of THEMIS 2 dust and evolved grains with $a_0=20$~nm with the function $\kappa_\nu=\kappa_0(\nu/\nu_0)^\beta$ between 160~$\mu$m and 2~mm, where $\kappa_0$ is adopted as $\kappa$ at 1~mm of the two respective dust models. The fitted $\beta$ values are 1.8 for THEMIS 2 dust and 2.2 for the evolved grains. Since the true $\kappa$ function varies with wavelength, the fitting results also vary, depending on the reference frequency $\nu_0$. For $\nu_0$ at 250~$\mu$m, for instance, the fitted $\beta$ values change to 2.1 for THEMIS 2 dust and 2.8 for the evolved grains. Nevertheless, the fitted $\beta$ of the evolved grains is always larger than the value of THEMIS 2 dust, independently of the choice of $\nu_0$.

\FloatBarrier  
\clearpage

\section{Results of radiative transfer modeling using only diffuse ISM type dust grains}\label{app:SOC_diffuse}

\FloatBarrier

\begin{figure}[t]
\centering
\includegraphics[width=9cm]{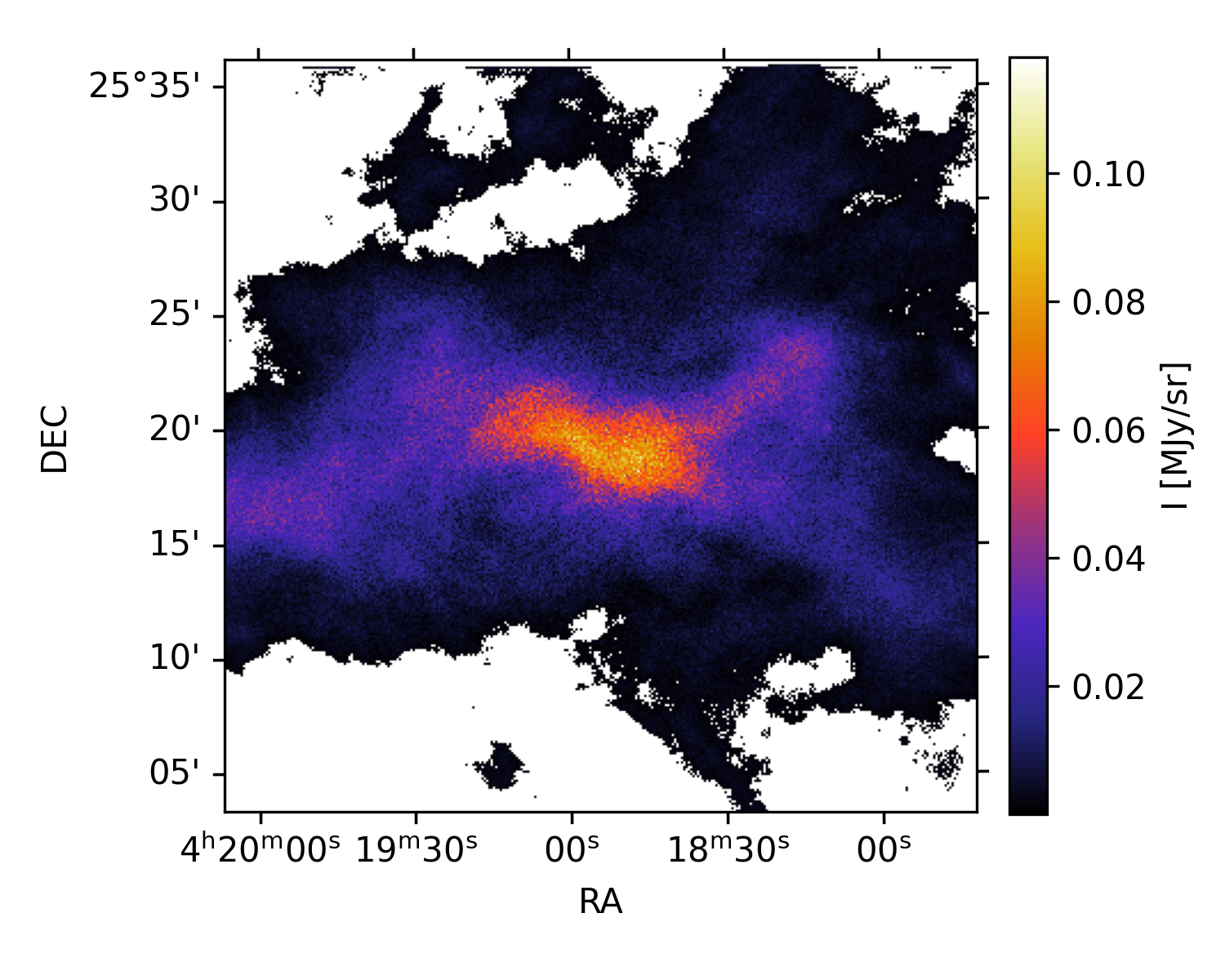}

\caption{Scattered light within \textit{Spitzer} IRAC1 3.6~$\mu$m filter modeled by SOC with input parameters of $A_\mathrm{V}=1$, $r_0=0.04$~pc, $\eta=1$ and diffuse type dust only.} 
\label{fig:SOC_scattering_diffuse}

\end{figure}

\begin{sidewaysfigure*}[htbp]
    \centering
    \includegraphics[width=20cm]{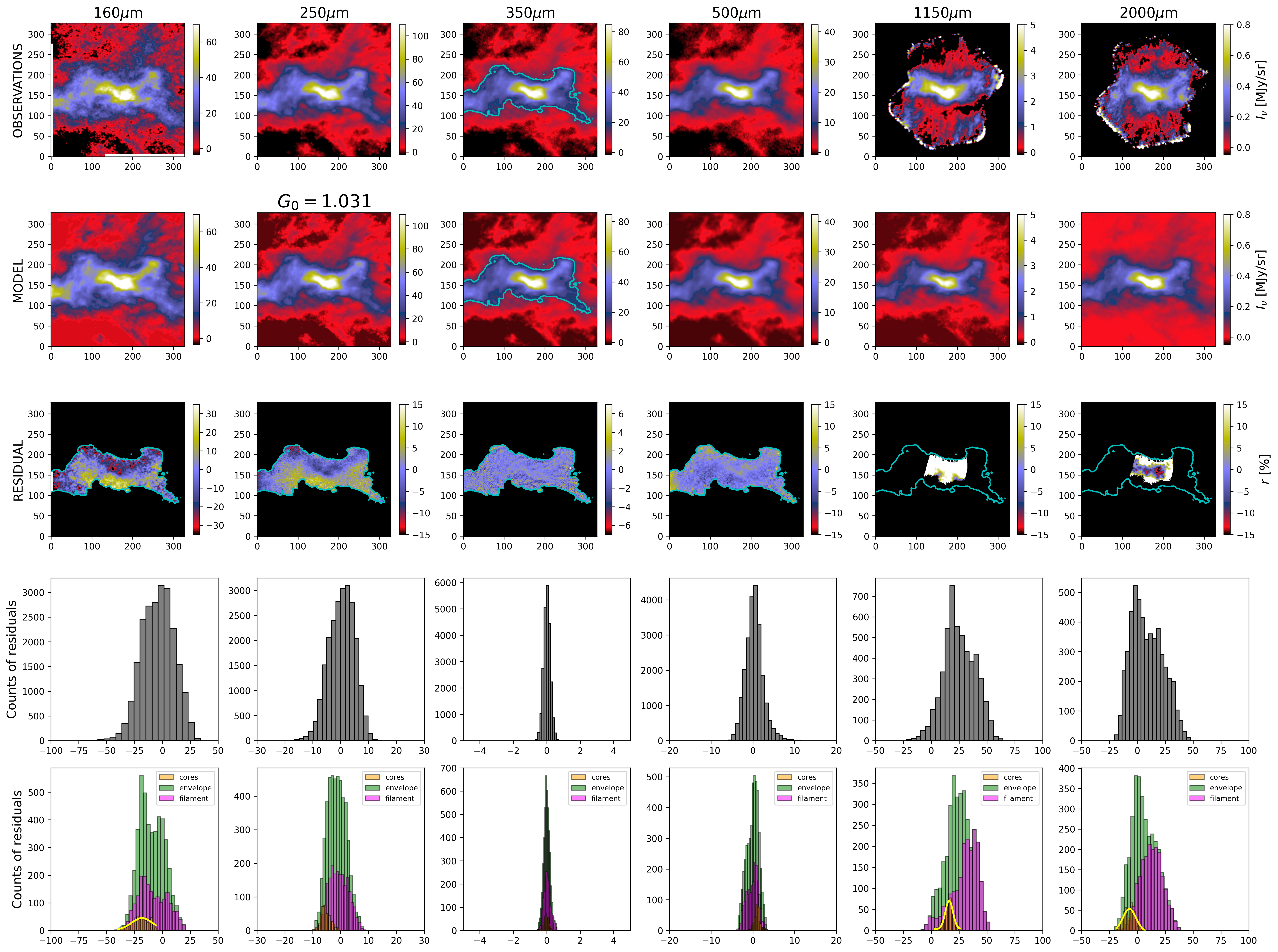}
    \caption{SOC modeling results of intensity maps with input parameters of $A_\mathrm{V}=1$, $r_0=0.04$~pc, $\eta=1$ and diffuse type dust only. Notations follow the definition of Fig.~\ref{fig:SOC_modeling_I}.} 
    \label{fig:SOC_modeling_I_diffuse}
\end{sidewaysfigure*}

The radiative transfer modeling results for selected parameter sets, with diffuse type dust only, are presented in Table~\ref{table:results_SOC_residuals_diffuse}. We also present one example of the modeled intensities and scattering with input parameters of $A_\mathrm{V}=1$, $r_0=0.04$~pc and $\eta=1$ in Figs.~\ref{fig:SOC_scattering_diffuse} and ~\ref{fig:SOC_modeling_I_diffuse}. The results obtained with a single dust model show that, as for models using two dust components, the parameters are degenerate. The peak position of the histogram of intensity residuals in the core region is $\lesssim20$\% at 160~$\mu$m and millimeter wavelengths for all combinations of input parameters. The absolute residuals are smaller than those obtained with two dust components at 160~$\mu$m, but larger at millimeter wavelengths. Moreover, the 2~mm residual of the cores is negative for all parameter sets, showing that the cores intensity is always overestimated at 2~mm in models using only diffuse type dust. Extinction is overestimated by a factor around two, slightly smaller than the factor of 2.8 found in the case of two dust populations. Because THEMIS 2 dust is more emissive than previous diffuse type dust models (Sect.~\ref{sec:dust_evolution}), it is possible to reproduce the emission and extinction using only diffuse type dust reasonably well by adjusting the cloud geometry. Compared to the results obtained with two dust populations, $r_{0, \parallel}$ and $r_{0, \perp}$ are larger ($\approx0.09$~pc for the former and up to 0.1~pc for the latter).  $\eta_{\parallel}$ and $\eta_{\perp}$ also increase ($\approx2.3$ for the former and up to 5 for the latter), probably to compensate the larger $r_0$.
However, for many combinations of input parameters, $n_c$ exceeds the value found with molecular line observation \citep[$n_\mathrm{H_2} < 5 \times 10^{4} \; \mathrm{cm}^{-3}$,][]{Pagani2010}. This larger value (compared to results with two dust populations) can be explained by the low FIR emissivity of diffuse type dust (about a factor of six smaller than evolved grains). 
For all parameter sets, max($I_\mathrm{sca}$) within the IRAC1 3.6~$\mu$m filter is about 0.1~MJy/sr, which is 2$\sigma$ larger than the observations. Thus, with only diffuse type dust, coreshine cannot be reproduced.

\begin{wraptable}{r}{15cm}
\caption{Results of radiative transfer modeling with diffuse type dust only for different values of the input parameters.}

\label{table:results_SOC_residuals_diffuse}

\setlength{\tabcolsep}{4pt}
\begin{tabular}{c|ccc|ccc|cc}  
     
$A_\mathrm{V}$ [mag] &1.0  & 1.0 & 1.0 & 1.0 & 1.0 & 1.0 & 1.0 & 1.0  \\
$r_0$ [pc] & 0.04 & 0.04 & 0.04 & 0.06  & 0.06 & 0.06 & 0.08 & 0.08 \\
$\eta$  &1.0 & 1.5 & 2.0  & 1.0 & 1.5 & 2.0 & 1.0 & 1.5 \\
\hline
\hline
$r_{0, \parallel}$ / $r_{0, \perp}$ [pc] & 0.09 / 0.07  & 0.09 / 0.05  & 0.08 / 0.04 & 0.09 / 0.09  & 0.09 / 0.08 & 0.08 / 0.07 & 0.09 / 0.10 & 0.09 / 0.10 \\
$\eta_\parallel$ / $\eta_\perp$ & 2.3 / 3.3  & 2.4 / 3.2 & 2.4 / 3.4  & 2.3 / 4.1  & 2.4 / 4.5 & 2.4 / 4.8 & 2.3 / 4.2 & 2.3 / 5.3 \\
$n_c$ [$10^4$ H$_2$/cm$^{3}$] & 4.7   & 8.8  & 14.5   & 3.5   & 5.9  & 9.0  & 2.8  & 4.5 \\

$G_0$ & 1.03 & 1.18 & 1.35 & 1.01 & 1.12 & 1.25 & 0.99 & 1.08\\

\hline
\hline

\shortstack{ core residual \\ 160~$\mu$m [\%]} & $-18.3$  $\pm 9.2$ & $-18.5$  $\pm8.8$ &  $-14.4$  $\pm 7.9$  & $-18.1$  $\pm 8.7$  & $-18.3$  $\pm 8.5$  & $-15.5$  $\pm 7.9$  & $-17.4$  $\pm 8.5$  & $-17.9$  $\pm 8.6$   \\ [3pt]

\shortstack{1.15~mm [\%]} & 16.2  $\pm 3.4$ &  $12.1$  $\pm 3.0$  & $-6.6$  $\pm 2.7$  & $16.8$  $\pm 3.4$  & $13.9$  $\pm 3.3$  & $10.0$  $\pm 2.9$  & $17.3$  $\pm 3.5$  &  $15.0$  $\pm 3.0$  \\ [3pt]

\shortstack{2~mm [\%]} & $-7.0$  $\pm 6.0$ &  $-13.6$  $\pm 6.8$  & $-22.2$  $\pm 8.3$  & $-6.0$  $\pm 6.0$  & $-10.6$  $\pm 6.6$  & $-16.6$  $\pm 6.9$ & $-5.3$  $\pm 6.0$  &  $-8.8$  $\pm 6.4$  \\ [3pt]

\hline
\hline

$k_{J}$  & 1.97 $\pm$ 0.02   &  2.15  $\pm 0.02$  & 2.36  $\pm$ 0.02  &  1.93  $\pm$ 0.02   &  2.06  $\pm$ 0.02  &  2.22 $\pm$ 0.02 &  1.91 $\pm$ 0.02 &  2.01  $\pm$ 0.02   \\ [3pt]

max($I_\mathrm{sca}$) [MJy/sr] & 0.10  & 0.10  & 0.10 & 0.10 & 0.11 & 0.11  &  0.10 & 0.10 \\

\end{tabular}
\tablefoot{Notations follow the definition in Table~\ref{table:results_SOC_residuals}.}
\end{wraptable}

\end{appendix}

\end{document}